\documentclass[prb,aps,twocolumn,showpacs,superscriptaddress]{revtex4-2}
\usepackage{amsfonts,amsmath,bm,multirow}
\usepackage[OT4]{fontenc}

\usepackage{graphicx}
\usepackage{xcolor}
\usepackage{epstopdf}
\usepackage{dsfont}
\usepackage{physics}
\usepackage{hyperref}
\usepackage{lipsum}
\usepackage{dcolumn}

\usepackage{mathtools}
\usepackage{hf-tikz}

\newcommand{\kp}{k\! \vdot\! p}
\newcommand{\rr}{\bm{r}}
\newcommand{\kk}{\bm{k}}

\definecolor{bred}{HTML}{e31a1c}
\definecolor{bgreen}{HTML}{33a02c}
\definecolor{bblue}{HTML}{1f78b4}

\definecolor{armygreen}{rgb}{0.29, 0.33, 0.13}
\definecolor{newred}{RGB}{255,70,70}
\definecolor{newcyan}{RGB}{0,200,255}

\renewcommand{\cp}{{\mathrm{c.p.}}}

\newcolumntype{L}{D{.}{.}{3,4}}

\begin{document}
	
	\title {Invariant expansion of the 30-band \texorpdfstring{$\bm{k}\!\vdot\!\bm{p}$}{p} model and its parameters for III-V compounds}
	
	\author{Krzysztof Gawarecki}
	\email{Krzysztof.Gawarecki@pwr.edu.pl}
	\affiliation{Department of Theoretical Physics, Wroc\l aw  University of Science and Technology, Wybrze\.ze Wyspia\'nskiego 27, 50-370 Wroc{\l}aw, Poland}
	\author{Pawe{\l} Scharoch}
	\affiliation{Department of Semiconductor Materials Engineering, Wroc\l aw  University of Science and Technology, Wybrze\.ze Wyspia\'nskiego 27, 50-370 Wroc{\l}aw, Poland}
	\author{Micha{\l} Wi{\'s}niewski}
	\affiliation{Department of Experimental Physics, Wroc\l aw  University of Science and Technology, Wybrze\.ze Wyspia\'nskiego 27, 50-370 Wroc{\l}aw, Poland}
	\author{Jakub Ziembicki}
	\affiliation{Department of Semiconductor Materials Engineering, Wroc\l aw  University of Science and Technology, Wybrze\.ze Wyspia\'nskiego 27, 50-370 Wroc{\l}aw, Poland}
	\author{Herbert S. M\k{a}czko}
	\affiliation{Department of Semiconductor Materials Engineering, Wroc\l aw  University of Science and Technology, Wybrze\.ze Wyspia\'nskiego 27, 50-370 Wroc{\l}aw, Poland}
	\author{Marta G{\l}adysiewicz}
	\affiliation{Department of Experimental Physics, Wroc\l aw  University of Science and Technology, Wybrze\.ze Wyspia\'nskiego 27, 50-370 Wroc{\l}aw, Poland}
	\author{Robert Kudrawiec}
	\affiliation{Department of Semiconductor Materials Engineering, Wroc\l aw  University of Science and Technology, Wybrze\.ze Wyspia\'nskiego 27, 50-370 Wroc{\l}aw, Poland}
	
	\begin{abstract}
		In this work we derive a ready-to-use symmetry invariant expansion of the full-zone $30$-band $\kp$ Hamiltonian for the $T_d$ point group.
		In order to find respective parameters, the band structures of III-V materials were calculated within a state-of-the-art Density Functional Theory (DFT) approach and used next as  targets to adjust the $\kp$ bands. A satisfactory agreement of the $\kp$ model with the DFT band structures, for all the tested zinc blende III-V semiconductors, has been achieved. Values of many of the parameters have not been known so far.
		We also compare the fitted $\kp$ parameters with the values calculated using momentum matrix elements obtained directly from the DFT. 
	\end{abstract}
	
	\maketitle
	
	\section{Introduction}
	\label{sec:intr}
	
	Zinc blende III-V semiconductors are key materials in modern optoelectronic devices including light emitting diodes (LED) and laser diodes (LDs) \cite{led2015,Stephan2016,Jung2017}. 
	The active part of the current LEDs and LDs are usually quantum wells (QW) or quantum dots (QD).
	Their size and content can be controlled precisely while growing with the use of the molecular beam epitaxy or the organometallic chemical vapor deposition methods \cite{Skierbiszewski2014,Bugajski2014,Yerino2017,Kuech2010}.
	A proper design of the structures in terms of the size and the content allows for an effective optimization of desired devices.
	For this purpose accurate band structure computations within efficient approaches are needed. A widely used class of the theoretical methods involves
	multiband $\kp$ models with the envelope function approximation \cite{Stier1999,Richard2004,Tomic2006,ElKurdi2010,Tomic2010,Gladysiewicz2013,Gladysiewicz2015,Ahmed2015,Campos2018}.
	However, some of their parameters are not well established for all the III-V semiconductors, especially when high-number-band $\kp$ models are considered.
	
	An accurate description of the first conduction bands across the entire Brillouin Zone (BZ) is of fundamental importance for quantitatively correct description of carrier transport phenomena \cite{Fischetti1988}, direct and indirect absorption calculations \cite{Ge2020}, optical gain calculations \cite{pascha}, etc.    
	Therefore $\kp$ models able to finely describe bands relevant for devices operation, with a complete set of material parameters, are desired.  
	Furthermore, some of III-V semiconductor compounds that are not completely parametrized (such as boron pnictides) 
	gained much attention recently in the field of optoelectronic devices \cite{Kudrawiec2020, Hidouri2020, Kudrawiec2019,El-Jaroudi2020}.
	
	There are three popular high-number-band $\kp$ models constructed with an aim to give an access to conduction bands across the entire BZ: 20-, 24-, and 30-band model.
	The 20-band model \cite{Pfeffer1996} is a direct extension of the 14-band model, giving a full access to the first conduction band with a satisfactory accuracy \cite{Cavassilas2001, Vogl1983}.
	However, it fails in description of the second conduction band, especially the 2$^{nd}$ L valley.
	The 24-band model is an extension of the 20-band model aiming at correcting the description of the valence bands and the two lowest conduction bands \cite{BenRadhia2003}.
	However, it still fails to finely reproduce the 2$^{nd}$ L valley and introduces a great number of parameters.
	These two models do not contain $d$ levels explicitly, which are essential for fine description of the two first conduction bands across the entire BZ.
	The 30-band $\kp$ model typically allows for such an accurate description \cite{Cardona1966, Richard2004}, and with a smaller number of parameters than in the 20- or 24-band models.
	The 30-band model is self-contained, which means it does not include additional bands via perturbations.
	For this reason, this model is much easier to use than other high-number-band models.
	Therefore, this model is chosen in our work for a description of all currently important III-V semiconductor compounds.
	In contrast to simple $\kp$ models, a lot of the parameters entering the high-number-band models are not directly accessible experimentally, so their determination is performed on the basis of fitting to band structures obtained within the state-of-the-art \emph{ab initio} approaches. Also, schemes of obtaining the $\kp$ parameters directly from the DFT calculations have been proposed~\cite{Jocic2020}.
	
	Currently, the best known methods of determining the band structure are based on  Density Functional Theory (DFT), either directly from Kohn-Sham equations with specially designed exchange-correlation (XC) potentials or within post-DFT methods (e.g. GW), in which Kohn-Sham solutions form a starting point \cite{GW}.
	However, the DFT methods are usually computationally heavy, and the results, like the wave functions, are given in  big data files which are inconvenient for using in standard programs for electronic devices modeling, many of which had been created before \emph{ab initio} methods were known.
	Therefore, the modeling usually exploits the $\kp$ method in approximations sufficient to represent the electronic structure in required range of the BZ, and the DFT methods are applied to benchmark the approximations and provide the required parameters.
	
	The multiband $\kp$ models contain a number of parameters, whose values depend on atoms forming a given material. However, a general structure of a $\kp$ model (including a non-zero pattern of its matrix elements) is determined by the symmetry. This universality is clearly visible in the theory of invariants, which allows to express the Hamiltonian by terms strictly connected to irreducible representations of a group describing the crystal symmetry~\cite{luttinger56,Trebin1979,LewYanVoon2009}. Comparing to a straightforward formulation, where the Hamiltonian is written as a single matrix (which is less practical for the $\kp$ models exceeding $8$ bands), the invariant expansion offers more compact formulas with a clear structure connected to the symmetry. Also, due to well defined multiplication relations between sub-matrices forming blocks of the Hamiltonian, it is much more convenient (compared to the explicit-matrix formulation) to use perturbation techniques (the L\"owdin partitioning)~\cite{Lowdin1950,LewYanVoon2009,Mielnik-Pyszczorski2018}. Invariant expansions for the eight- and fourteen-band $\kp$ models for the $T_\mathrm{d}$ point group were derived decades ago and well established~\cite{Suzuki1974,Trebin1979,mayer91,LewYanVoon2009,Winkler2003}. Recently, in Ref.~\cite{PhysRevB.104.085137} authors used symmetry considerations to describe a way to construct any $\kp$ model for any symmetry, and they give a very large database containing representation matrices (4 857 832  matrix blocks). However, its application is not straightforward.
	In consequence, although the properties of arbitrary-size models for the $T_\mathrm{d}$ point group are known~\cite{Wanner2017,PhysRevB.104.085137}, no convenient ready-to-use invariant-based formulation has been provided for the $30$-band model so far.
	
	In this paper, we derive an explicit form of the Hamiltonian given in terms of the symmetry invariants. We also offer the parameters for a wide class of the III-V semiconductor materials found using the DFT (\emph{ab initio}) reference electronic structures. The 30-band $\kp$ model used here is usually sufficient to represent the electronic bands in  the full BZ which is often necessary to properly describe the properties connected with the indirect electronic transitions. Finally, in the case of GaAs, we compare the fitted parameters to the values extracted directly from momentum matrix elements calculated in the DFT approach. We show a very good agreement for most of the parameters.
	
	The paper is organized as follows.  In Sec.~\ref{sec:kp30}, we describe the $30$-band $\kp$ model and its invariant expansion. In Sec.~\ref{sec:dft}, we present the DFT approach.
	Sec.~\ref{sec:extr} is devoted to the parameters related to interband momentum matrix elements, which are further obtained directly from the DFT calculations.
	In Sec.~\ref{sec:fit} we briefly describe the $\kp$ to DFT fitting procedure. In Sec.~\ref{sec:results}, we present examples of the $\kp$ vs DFT band structures and the values of evaluated $\kp$ parameters. In Sec.~\ref{sec:examples}, we present an exemplary application of the model. Sec.~\ref{sec:concl} contains a summary. In Appendix~\ref{app:ham}, one can find an explicit form of the Hamiltonian and matrices including a detailed description of derivation. Appendix~\ref{app:reduction} describes the reduction from the 30-band $\kp$ to models with fewer bands.  In Appendix~\ref{app:mel}, the calculations related to the interband momentum matrix elements are presented. Finally, the complete set of the plotted band structures is given in the attached Supplementary Material~\cite{Supplementary}.

	\section{The 30 \texorpdfstring{$\bm{k}\!\vdot\!\bm{p}$}{p} model}
	\label{sec:kp30}
	In the approach of $30$-band $\kp$ applied to materials of the diamond structure (the $O_\mathrm{h}$ point group), the only non-vanishing interband momentum matrix elements are between the basis states of the different parity with respect to the inversion~\cite{Richard2004}. In the case of the zinc blende atomic structure, the inversion symmetry is broken~\cite{Bir1974} and number of additional parameters have to be introduced to the model~\cite{Richard2004}. However, their impact on the main spectral properties is usually weaker compared to the parameters non-vanishing for the diamond structure~\cite{Winkler2003}. On the other hand, the bulk inversion asymmetry is important for spin-related properties (which is manifested via the Dresselhaus coupling)~\cite{Dresselhaus1955}. Since such effects in many applications are considered only in the vicinity of the $\Gamma$ point, near the fundamental band gap, we keep the parameters related to the inversion asymmetry in the part of the Hamiltonian corresponding to the $14$-band $\kp$ model~\cite{mayer91,Winkler2003} and neglect them otherwise~\cite{Richard2004} (except to one additional parameter in the valence band)  . 
	
	The basis used in the considered model contains $30$ states which can be grouped according to the irreducible representations ($\Gamma_6$, $\Gamma_8$ or $\Gamma_7$) of the $\mathrm{T}_\mathrm{d}$ point group~\cite{Richard2004}. Then, the Hamiltonian $H$ can be decomposed into the blocks
	\begin{equation}
		H_{n \alpha n' \alpha'} = \mel{\bm{\Gamma}_{n\alpha}}{H}{\bm{\Gamma}_{n'\alpha'}},
	\end{equation}
	where $\ket{\bm{\Gamma}_{n\alpha}}$ are pseudospinors with $n = 6,7,8$ denoting the representation and $\alpha = $``w", ``v", ``c", ``u", ``t", ``d", ``q" labeling the band set.
	Size of matrices corresponding to each block are determined by $\dim{\Gamma_n} \times \dim{\Gamma_{n'}}$~\cite{Winkler2003}. Such blocks within the theory of invariants can be further expressed as combinations of irreducible tensor components~\cite{Bir1974,Winkler2003,Wanner2017}.
	
	The block-diagonal parts of the Hamiltonian can be written as 
	\begingroup
	\allowdisplaybreaks
	\begin{align*}
		H_{\mathrm{6\alpha6\alpha}} &=  \qty( \widetilde{E}_{\mathrm{6\alpha}} + \frac{\hbar^2}{2m_0} k^2 ) \; \mathbb{I}_2 ,\\
		H_{\mathrm{8\alpha8\alpha}} &=  \qty( \widetilde{E}_{\mathrm{8\alpha}} + \frac{1}{3} \widetilde{\Delta}_{\alpha\alpha} + \frac{\hbar^2}{2m_0} k^2 ) \; \mathbb{I}_4 ,\\
		H_{\mathrm{7\alpha7\alpha}} &=  \qty( \widetilde{E}_{\mathrm{7\alpha}} -\frac{2}{3} \widetilde{\Delta}_{\alpha\alpha}+ \frac{\hbar^2}{2m_0} k^2 ) \; \mathbb{I}_2,
	\end{align*}
	\endgroup
	where $\mathbb{I}_n$ is the $n$-th size identity matrix, $\widetilde{E}_{\mathrm{m\alpha}}$ are bare energies at $\kk=0$ without the spin-orbit coupling, and $m_0$ is the free electron mass. Finally, the diagonal corrections due to the spin-orbit coupling are introduced via the parameters $\widetilde{\Delta}_{\alpha\alpha}$. As the energies $\widetilde{E}_{\mathrm{m\alpha}}$ do not contain the spin-orbit part, it is convenient to label them using the single-group notation~\cite{Richard2004} (with the representations $\Gamma_1$, $\Gamma_5$ and $\Gamma_3$), where $\widetilde{E}_{\mathrm{6\alpha}} = {E}_{\mathrm{1\alpha}}$ and
	$\widetilde{E}_{\mathrm{8\alpha}} = \widetilde{E}_{\mathrm{7\alpha}} = {E}_{\mathrm{5\alpha}} $ (except $\alpha = $``t", where $\widetilde{E}_{\mathrm{8t}} = {E}_{\mathrm{3t}}$). 
	The off-diagonal blocks of the Hamiltonian can be written in a general form
	\begingroup
	\allowdisplaybreaks
	\begin{align*}
		H_{\mathrm{6 \alpha 8 \beta}} &=  \sqrt{3} \mathcal{P}_{\alpha \beta} \qty(T_x k_x + \cp),\\
		H_{\mathrm{6 \alpha 7 \beta}} &=  -\frac{1}{\sqrt{3}} \mathcal{P}_{\alpha \beta} \qty(\sigma_x k_x + \cp),\\
		H_{\mathrm{8 \alpha 8 \beta}} &=  -\frac{2}{3} \mathcal{Q}_{\alpha \beta} \qty(\{J_y,J_z\} k_x + \cp) \\ & \quad - \sqrt{30} \mathcal{R}_{\alpha\beta} (D_x k_x + \cp) + \frac{1}{3} \widetilde{\Delta}_{\alpha \beta} \mathbb{I}_4,\\
		H_{\mathrm{8 \alpha 7 \beta}} &= -2 \mathcal{Q}_{\alpha \beta} (T^\dagger_{yz} k_x+\cp) \\ & \quad +\sqrt{6} \mathcal{R}_{\alpha\beta} (T^\dagger_x k_x + \cp),\\
		H_{\mathrm{7 \alpha 7 \beta}} &= -\frac{2}{3} \widetilde{\Delta}_{\alpha \beta} \mathbb{I}_2,
	\end{align*}
	\endgroup
	where $\mathcal{P}_{\alpha \beta}$, $\mathcal{Q}_{\alpha \beta}$, and $\mathcal{R}_{\alpha\beta}$ are parameters related to the interband momentum matrix elements, $\{A,B\} = (AB+BA)/2$,\, $T_i$ are $2 \times 4$ matrices connecting different blocks, $J_i$ are $4\times4$ matrices related to the total angular momentum, $T_{ij} = T_i J_j + T_j J_i$, and $D_i$ are $4 \times 4$ matrices. While we took the $T_i$, $J_i$, and $T_{ij}$ from the literature~\cite{Trebin1979}, we derived the $D_i$ matrices from the Wigner-Eckart theorem. The remaining blocks can be calculated from the hermitian conjugates $H_{\mathrm{n \alpha m \beta}} = H^\dagger_{\mathrm{m \beta n \alpha}}$. The calculation details, explicit definitions of the basis functions, matrices, and a ready-to-use form of the Hamiltonian are given in Appendix~\ref{app:ham}. The invariant expansions of the Hamiltonian blocks, except to the ones involving ``8t", have the well known structure of the $14$-band $\kp$ model~\cite{Trebin1979,Winkler2003} with the same matrices $T_i$, $J_i$, $T_{ij}$; yet more parameters $\mathcal{P}_{\alpha \beta}$, $\mathcal{Q}_{\alpha \beta}$ are needed.
	
	In this work, to model the properties of the III-V materials, we keep the nonzero parameters enlisted in Table~\ref{tab:nonzero}. As they impact on the bands of interest is expected to be rather subtle, we neglected off-diagonal spin-orbital parameters,
	except to $\Delta^-$. 
	\begin{table}
		\caption{\label{tab:nonzero}List of nonzero parameters, taken into account in the fitting procedure. We assume $E_{\mathrm{8v}} = {E}_{\mathrm{5v}} + (1/3)\Delta_\mathrm{v} = 0$ as the reference energy (the top of the valence band). }
		\begin{ruledtabular}
			\begin{tabular}{l l l l}
				parameters & blocks &  abbrev. & note \\[0.06in]
				\hline
				$\mathcal{P}_{\mathrm{cv}}$ & $H_\mathrm{6c8v}$, $H_\mathrm{6c7v}$ & $P_0$ &  \\
				$\mathcal{P}_{\mathrm{cd}}$ & $H_\mathrm{6c8d}$, $H_\mathrm{6c7d}$ & $P_1$ &  \\
				$\mathcal{P}_{\mathrm{qv}}$ & $H_\mathrm{6q8v}$, $H_\mathrm{6q7v}$ & $P_2$ &  \\
				$\mathcal{P}_{\mathrm{qd}}$ & $H_\mathrm{6q8d}$, $H_\mathrm{6q7d}$ & $P_3$ &  \\
				$\mathcal{P}_{\mathrm{uc}}$ & $H_\mathrm{6u8c}$, $H_\mathrm{6u7c}$ & $P_4$ &  \\
				$\mathcal{P}_{\mathrm{wc}}$ & $H_\mathrm{6w8c}$, $H_\mathrm{6w7c}$ & $P_5$ &  \\
				$\mathcal{P}_{\mathrm{cc}}$ & $H_\mathrm{6c8c}$, $H_\mathrm{6c7c}$ & $P'_0$ & inv. asymm. \\
				$\mathcal{P}_{\mathrm{wv}}$ & $H_\mathrm{6w8v}$, $H_\mathrm{6w7v}$ & $P'_1$ & inv. asymm. \\
				$\mathcal{Q}_{\mathrm{cv}}$ & $H_\mathrm{8c8v}$, $H_\mathrm{8c7v}$, $H_\mathrm{7c8v}$ & $Q_0$ &  \\
				$\mathcal{Q}_{\mathrm{dc}}$ & $H_\mathrm{8d8c}$, $H_\mathrm{8d7c}$, $H_\mathrm{7d8c}$ & $Q_1$ &  \\
				$\mathcal{R}_{\mathrm{tv}}$ & $H_\mathrm{8t8v}$, $H_\mathrm{8t7v}$ & $R_0$ &  \\
				$\mathcal{R}_{\mathrm{td}}$ & $H_\mathrm{8t8d}$, $H_\mathrm{8t7d}$ & $R_1$ &  \\
				$\widetilde{\Delta}_{\mathrm{vv}}$ & $H_\mathrm{7v7v}$, $H_\mathrm{8v8v}$ & $\Delta_\mathrm{v}$ &  \\
				$\widetilde{\Delta}_{\mathrm{cc}}$ & $H_\mathrm{7c7c}$, $H_\mathrm{8c8c}$ & $\Delta_\mathrm{c}$ &  \\
				$\widetilde{\Delta}_{\mathrm{dd}}$ & $H_\mathrm{7d7d}$, $H_\mathrm{8d8d}$ & $\Delta_\mathrm{d}$ &  \\
				$\widetilde{\Delta}_{\mathrm{cv}}$ & $H_\mathrm{7c7v}$, $H_\mathrm{8c8v}$ & $\Delta^-$ & inv. asymm. \\
			\end{tabular}
		\end{ruledtabular}
	\end{table}

	\section{The \emph{ab initio} approach}
	\label{sec:dft}
	The main purpose of the DFT calculations in this work was to obtain high quality band structures which are needed as a reference to adjust the $30$-band $\kp$ model. The method commonly regarded as the best one to represent the band structure is the DFT-based GW approach \cite{GW1,GW2}. However, in this work, we decided to use the HSE hybrid functional~\cite{HSE} which is known to provide, at a relatively low computational cost,  an accurate description of electronic bands, as opposed to the standard LDA and GGA schemes, where the well known band gap problem is exhibited. Moreover, in the hybrid functional, the results may be further improved by fitting band gaps to the experimental ones with the $\mu$ screening parameter~\cite{kim_marsman_kresse_tran_blaha_2010}. For all the materials considered in this work (except BSb, as in this case, the experimental band gap is unknown) the value of the $\mu$ parameter was rescaled to yield experimental band gaps~\cite{vurgaftman_meyer_ram-mohan_2001,schormann_2006,thompson_auner_zheleva_jones_simko_hilfiker_2001,BAs_BG,BN_BG,BP_BG}. The calculations of band structures were made with VASP code and PAW method~\cite{VASP,VASP_PAW}. The Brillouin zone was sampled with the 11x11x11 grid and the energy cutoff was set to 600 eV. 
	
	The lattice parameters were optimized with the use of ABINIT~\cite{Abinit} package together with the PAW method~\cite{Abinit_PAW} and the Wu and Cohen GGA functional~\cite{wu_cohen_2006}. In Ref.~\cite{haas_tran_blaha_2009}, the authors showed that this functional yields very accurate lattice constants for this group of materials. Our calculations confirm this.  The mean absolute deviation of their values with respect to experiment is about 0.3 \%. In VASP, this particular functional is not implemented.

	\section{Extracting the \texorpdfstring{$\bm{k}\!\vdot\!\bm{p}$}{p} model parameters directly from the DFT calculations}
	\label{sec:extr}
	The parameters that enter the 30-band $\kp$ Hamiltonian can be divided into three classes: the energies $\widetilde{E}_{\mathrm{m\alpha}}$, the spin-orbit coupling parameters $\widetilde{\Delta}_{\alpha\beta}$, and the $\mathcal{P}_{\alpha \beta}$, $\mathcal{Q}_{\alpha \beta}$, $\mathcal{R}_{\alpha\beta}$ which are related to interband momentum matrix elements.
	The $\widetilde{E}_{\mathrm{m\alpha}}$ and the diagonal $\widetilde{\Delta}_{\alpha\alpha}$ are directly read from the DFT band energies at the $\Gamma$ point (which is an approximation, because the band energies at this point are also affected by the off-diagonal $\widetilde{\Delta}_{\alpha\beta}$~\cite{Cardona1988}).
	As the $\kp$ basis is finite, the parameters related to interband momentum matrix elements need to be rescaled to obtain a better agreement to the target data. 
	The $30$-band $\kp$ contains a relatively large number of basis states, hence one can expect that bare $\mathcal{P}_{\alpha \beta}$, $\mathcal{Q}_{\alpha \beta}$, $\mathcal{R}_{\alpha\beta}$, calculated from the momentum matrix elements within the DFT approach, could give a reasonable approximation (or at least an initial value for further fitting).
	
	While the parameters, which are given in the further part of the paper comes from the fitting, in the case of GaAs we also extract them directly from the DFT, and compare the values obtained in both methods.
	The VASP code contains a procedure, allowing to get the momentum matrix elements. With this, we calculate $(p_{i})_{nm} = \mel{\Psi_n}{\hat{p}_{i}}{\Psi_m}$ for the states $\ket{\Psi_l}$ at the $\Gamma$ point. To obtain the values of $\mathcal{P}_{\alpha \beta}$, $\mathcal{Q}_{\alpha \beta}$, and $\mathcal{R}_{\alpha\beta}$, one needs to extract the matrix elements between the basis states forming $\ket{\Psi_l}$. This, however, requires a caution due to degeneracies and complex phases. 
	A detailed description of the applied procedure is presented in the Appendix~\ref{app:mel}.
	
	\section{Fitting the 30 band \texorpdfstring{$\bm{k}\!\vdot\!\bm{p}$}{p} model parameters}
	\label{sec:fit}
	\begin{figure*}[t]
		\begin{center}
			\includegraphics[width=130mm]{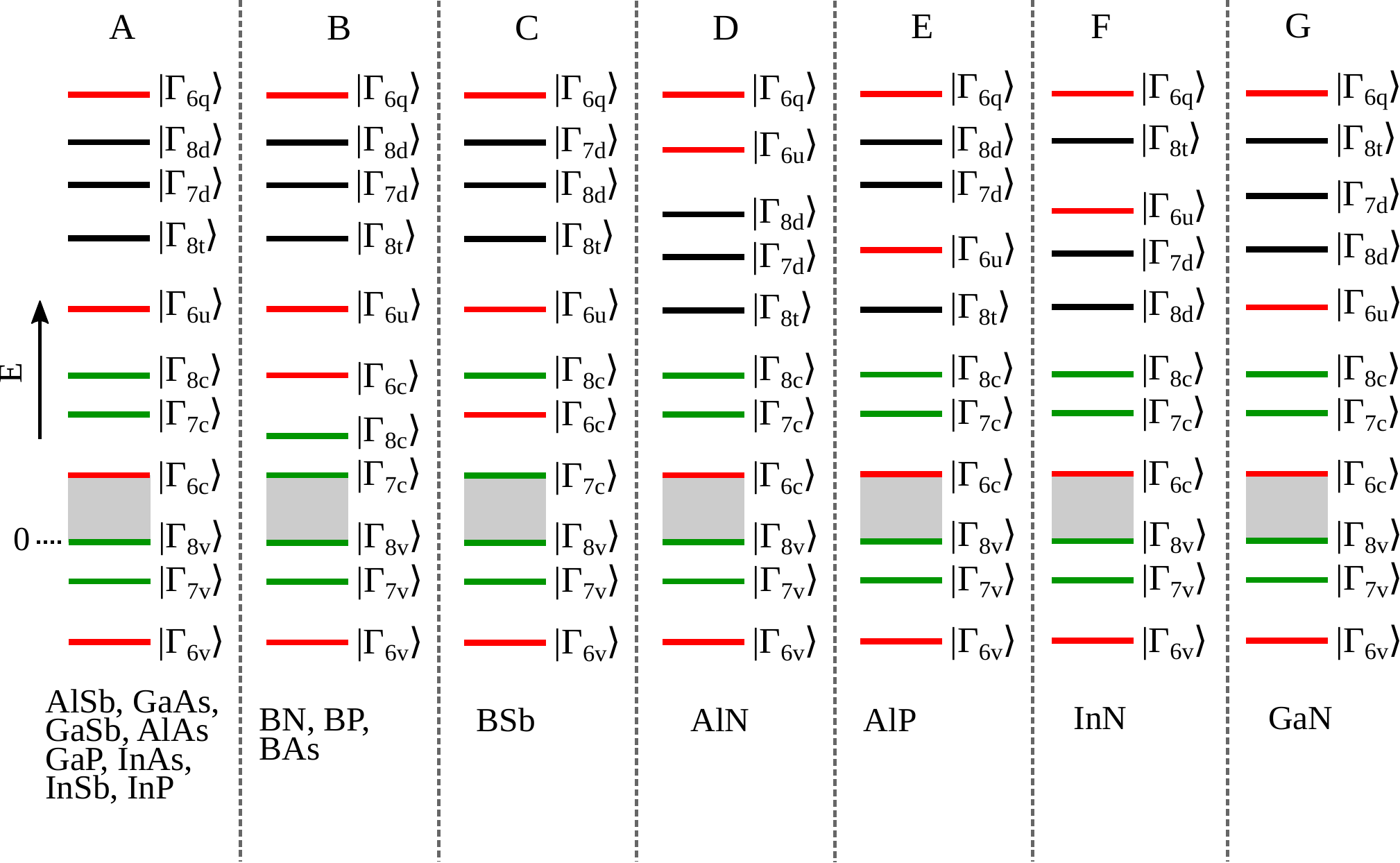}
		\end{center}
		\caption{\label{fig:bulk}\textcolor{gray}({Color online) Schematic band alignments at the $\Gamma$ point, for various III-V semiconductor compounds. The color indicates the dominant contribution from the atomic orbitals: red is the $s$-type, green is the $p$-type, and black denotes the $d$-type. We assume the energy $E = 0$ at the top of the valence band. The gray box indicates the fundamental energy gap.} }
	\end{figure*}
	The procedure of finding the $30$ band $\bm{k}\cdot\bm{p}$ parameters, by adjusting the $\bm{k}\cdot\bm{p}$ band structure to the DFT reference one, has been described in details in Ref.~\cite{pascha}. A central point of this procedure is a minimization of the objective function $S$, defined as:

	\begin{equation}\label{Eq:S}
		S=\sqrt{\frac{1}{N}\sum_{i=1}^N\left(E_{\mathrm{DFT},i}-E_{\bm{k}\cdot\bm{p},i}\right)^2},
	\end{equation}
	where $E_{\mathrm{DFT},i}$ are energies of the DFT bands in a number of points in the BZ. In this work $45$ points in  along high symmetry lines $X-\Gamma-L$ in the BZ have been chosen. The number of points is higher than in Ref.~\cite{pascha} because of the lower crystal symmetry (zinc blende vs diamond).  The $E_{\bm{k}\cdot\bm{p},i}$ are the corresponding energies of the fitted $\bm{k}\cdot\bm{p}$ band structure. The summation runs over the $\bm{k}$ points in the BZ and over the bands, thus $i$ is a compound index. The function, after minimization, is also a measure of the accuracy of the fit, with zero representing a perfect match. The method of steepest descent has been used to find a minimum, which was previously shown to work very well for diamond structure materials~\cite{pascha}. The diagonal $\widetilde{E}_{\mathrm{m\alpha}}$ and  $\widetilde{\Delta}_{\alpha\alpha}$ parameters are taken from the $\Gamma$-point energies~\cite{pascha}, while the off-diagonal $\Delta^-$ is fitted together with the other parameters. This procedure is a kind of approximation, because the energies at the $\Gamma$ point are already affected by the off-diagonal $\Delta^-$ term~\cite{Cardona1988}.

	It should be pointed out that, in the fitting procedure, a proper assignment of the $E_{\bm{k}\cdot\bm{p}}$ bands to the DFT ones is crucial. As it is a known fact, the order of bands, classified by group representations, is not the same in all compounds. The identification of the bands in the $\Gamma$ point has been done with use of DFT tools (like projection of states on atomic orbitals, etc.), and the results are shown in Fig.\ref{fig:bulk}.  It can be seen that there are $7$ different cases of the band order, among which the most common is the one represented by GaAs.

	Fitting procedures similar to the one described above have already been used by other authors, e.g. \cite{Bastos, Faria}, although there are some differences. In the cited works, a lower number of bands has been included in the $\bm{k}\cdot\bm{p}$ ($6$ and $8$) which allowed to apply a partly analytic form (the secular equation) and which caused the necessity of testing the values of the parameters as a function of the range of $k$ taken into account. We expect a good match in the whole BZ for the $30$-band $\bm{k}\cdot\bm{p}$, according to our previous experience, and thus, the $k$-points range testing was not necessary and the points have been chosen arbitrarily. Also, a purely numerical diagonalization of the $\bm{k}\cdot\bm{p}$ Hamiltonian has been applied for the fitting.
	We note that the fitting procedure, as well as the calculations described in the previous section, do not give reliable information about the signs of the parameters that enter the off-diagonal Hamiltonian blocks. 
	Therefore, there is a some uncertainty about the signs of ${P}_{i}$, ${P}_{i}'$ ${Q}_{i}$, ${R}_{i}$ and $\Delta^-$, which is a drawback of our sets of parameters.
	We chose the most likely signs of the parameters not vanishing at the inversion symmetry by inspecting the wave functions obtained from the DFT. However, due to the uncertainty, we advise a caution in interpolation of the parameters for alloys.

	\section{Results and discussion}
	\label{sec:results}
	
	We begin the presentation and discussion of the results from GaAs, which is a well known III-V semiconductor. The  $\bm{k}\cdot\bm{p}$ parameters for this compound can be easily found in the literature~\cite{vurgaftman_meyer_ram-mohan_2001, vurg2, Bastos, Shok, Bouj, Waqas, Ramos}.
	However, most of the available parameters
	correspond to the $8$ band  $\bm{k}\cdot\bm{p}$ model,
	derived from experiment (e.g. Ref.~\cite{Shok}), from a fitting to known band structures (e.g. Ref.~\cite{Bastos}), or based on an extended literature review \cite{vurgaftman_meyer_ram-mohan_2001, vurg2}. In this work we present the parameters for the $30$ band $\bm{k}\cdot\bm{p}$ model, whose advantage is an adequate representation of band structure in the whole BZ, as it has been shown in Ref.~\cite{pascha}. 
	
\begin{figure}[h!]
		\begin{center}
			\includegraphics[width=.45\textwidth, trim={4 1.5cm 4 1.5cm}]{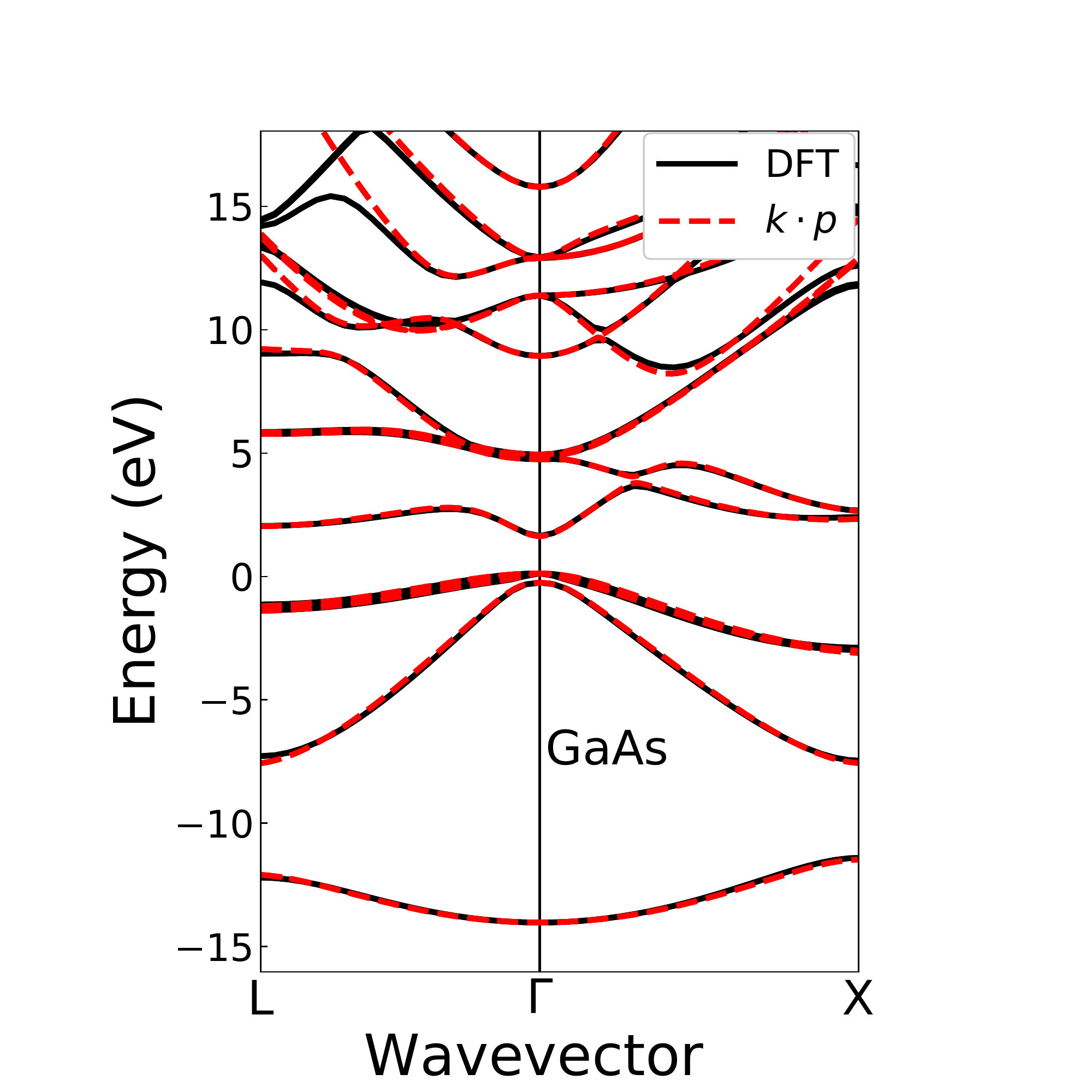}
		\end{center}
		\caption{\label{fig:GaAs}\textcolor{black}{Comparison of the DFT and the $\kp$ band structures for GaAs.} }
	\end{figure}

	Figure \ref{fig:GaAs} shows the GaAs band structures, the $30$-band $\bm{k}\cdot\bm{p}$ one together with the DFT one used as a reference to find the $\kp$ parameters. As it can be seen, the agreement is very satisfactory, particularly in $6$ highest valence bands and $8$ lowest conduction bands. This is a sufficient representation to model semiconductor devices, like LEDs, lasers, or solar cells. The respective parameters are given in the appropriate table later in the article.
	
\begin{figure}[tb]
		\begin{center}
			\includegraphics[width=.48\textwidth, trim={2 6.5cm 2 6.5cm}]{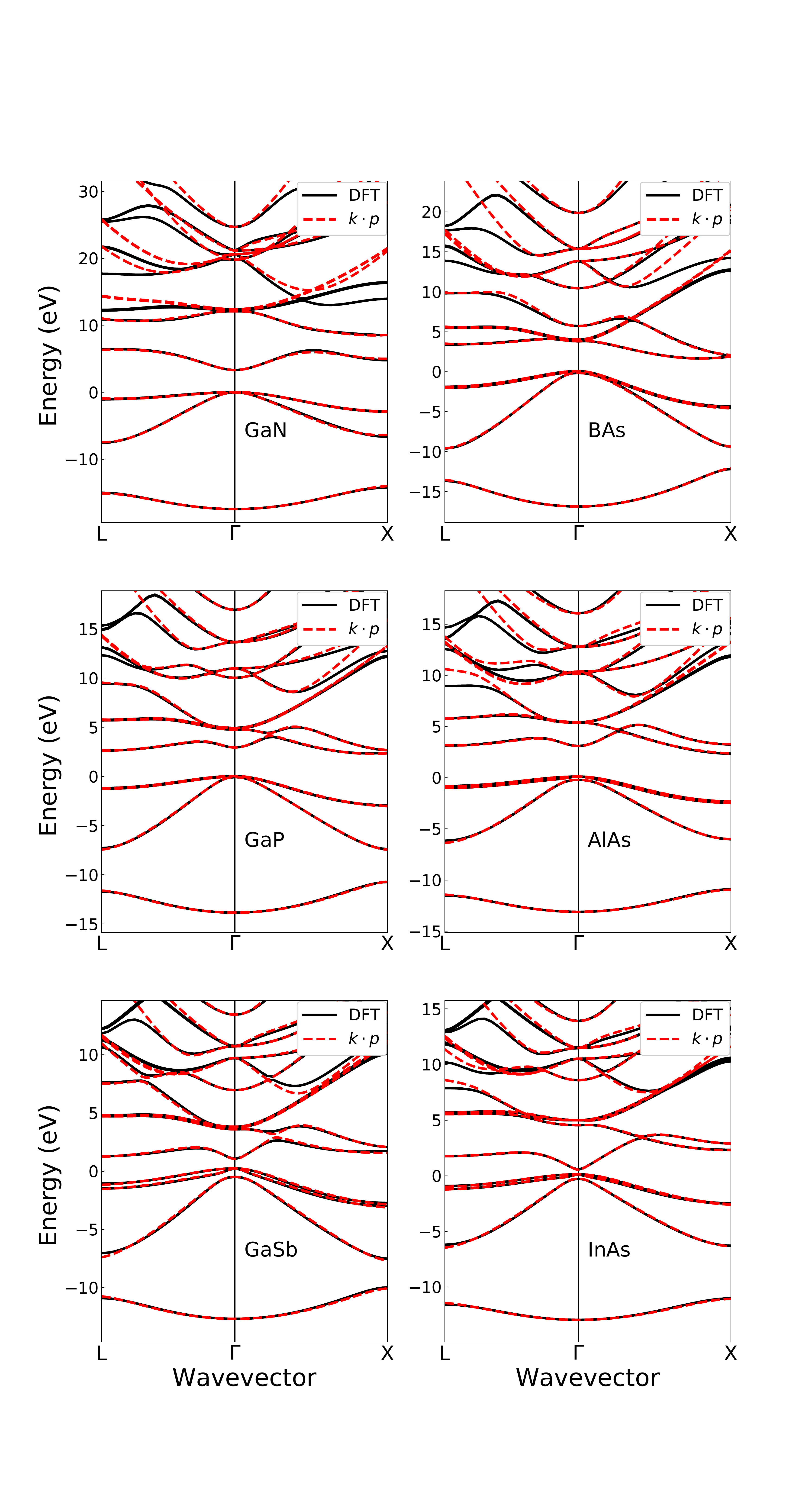}
		\end{center}
		\caption{\label{fig:GaX_YAs}\textcolor{black}{Comparison of the DFT and the $\kp$ band structures for Ga$X$ (left) and $Y$As (right) materials.} }
	\end{figure}

	The GaAs can be treated as a kind of a central compound among the III-V systems. In this work, we have calculated the $\kp$ parameters for sixteen III-V semiconductor binary compounds, but we consider six compounds related to GaAs in order to have a look at the chemical trends, namely Ga$X$ and $Y$As, where $Y$ and $X$ stand for the group III and V elements, respectively. The respective graphs are shown in Fig.~\ref{fig:GaX_YAs}. The overall view confirms the conclusion related to GaAs: the agreement between the DFT and the $30$-band $\bm{k}\cdot\bm{p}$ band structures is good, particularly in the regions significant for the device modeling. The indirect gap is very well reproduced for GaP in Ga$X$ as well as for BAs and AlAs in $Y$As. This proves the reliability of the applied procedure.
	For the remaining nine compounds (the figures are in the attached Supplementary Material~\cite{Supplementary}), the electronic structure obtained by the DFT method is also well reconstructed using the $30$-band $\bm{k}\cdot\bm{p}$ model.
	
	The values of the $\kp$ parameters for the sixteen III-V semiconductor compounds are listed in Tabs.~\ref{tab:BX}-\ref{tab:InX}. They can be directly used for physical phenomena analysis or in device modeling. As demonstrated in Appendix~\ref{app:reduction}, they could be also used for derivation of the standard parameters (the effective masses, the Luttinger parameters, etc.) for fewer-band models describing the vicinity of the $\Gamma$ point.
	However, it should be recalled that our parameters are optimized for the full BZ rather than for any specific $\kk$ point. Therefore, more accurate parameters for the models with smaller number of bands could be obtained by a fitting in a narrower part of the BZ. 
	
	\begin{figure}[tb]
		\begin{center}
			\includegraphics[width=.48\textwidth, trim={2 6.5cm 2 6.5cm}]{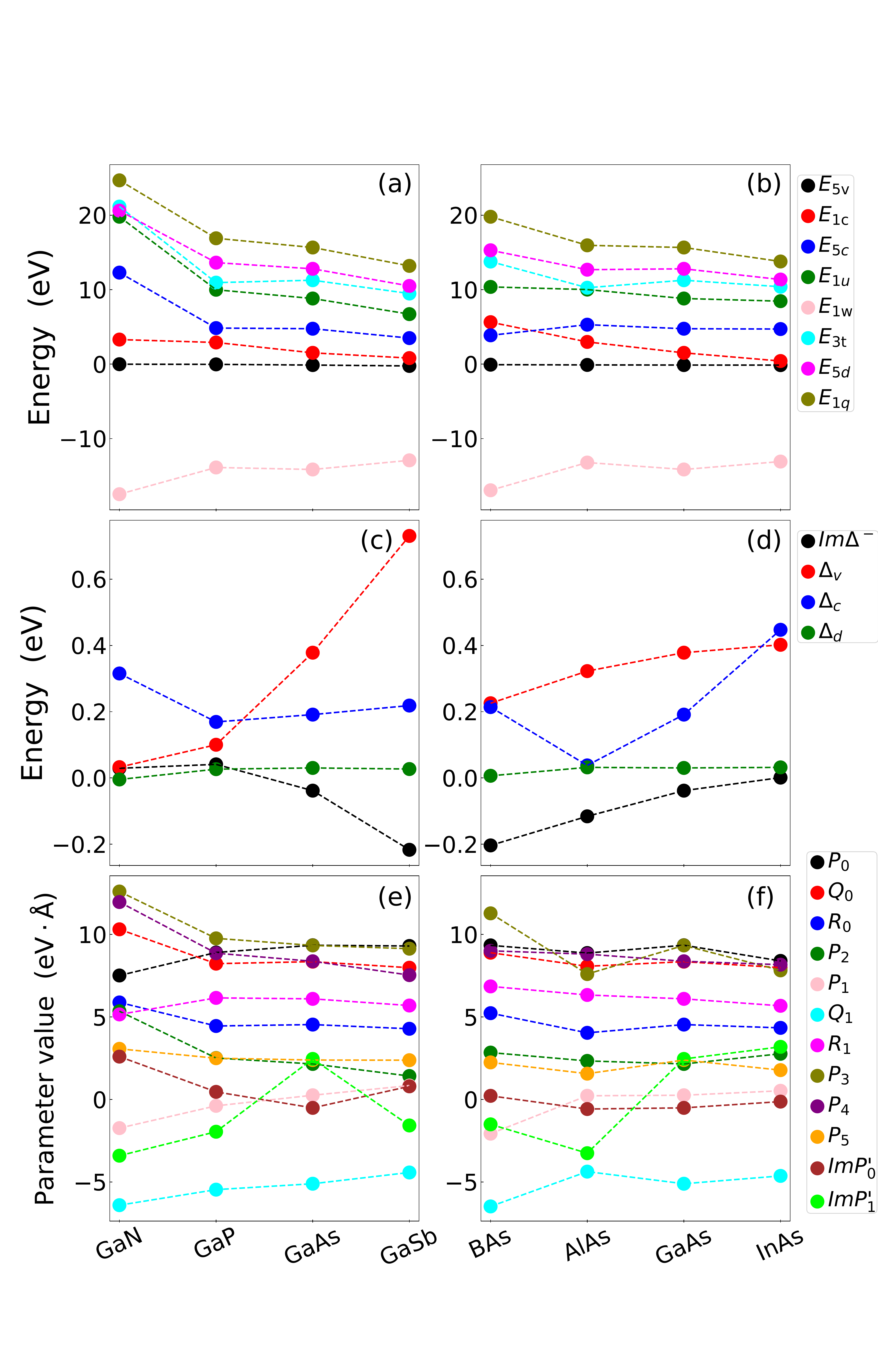}
		\end{center}
		\caption{\label{fig:all_GaAs}\textcolor{black}{Parameters for: Ga$X$ compounds (a,c,e) and $Y$As compounds  (b,d,f).} }
	\end{figure}
	The chemical trends in the $\kp$ parameters can be observed in Figs.~\ref{fig:all_GaAs}(a-f), where the Ga$X$ and $Y$As
	values of the energies (a,b), the spin-orbit coupling parameters (c,d), and the parameters related to momentum matrix elements (e,f) are visualised. The Ga$X$ ($Y$As) compounds are ordered with respect to the atomic number of $X$ ($Y$).
	In many cases the behavior of parameters is monotonic and the variation is rather weak, but the chemical trends are clearly seen. For the energy parameters (a,b), it should be recalled that according to the applied convention the \emph{ab initio} reference (zero) energy level is the top of the valence band.

	\renewcommand{\arraystretch}{1.2}
	\begin{table}
		\caption{\label{tab:BX}Parameters for B$X$ materials}
		\begin{ruledtabular}
			\begin{tabular}{lLLLL}
				& \multicolumn{1}{c}{BN} & \multicolumn{1}{c}{BP} & \multicolumn{1}{c}{BAs} & \multicolumn{1}{c}{BSb} \\
				\hline
				$E_{\mathrm{g}}$\hfill(eV) & 6.595 & 1.913 & 1.571 & 1.113 \\
				\hline
				$E_{\mathrm{1w}}$\hfill(eV) & -22.252 & -17.203 & -16.942 & -15.215 \\
				$E_{\mathrm{5v}}$\hfill(eV) & -0.008 & -0.015 & -0.075 & -0.121 \\
				$E_{\mathrm{1c}}$\hfill(eV) & 12.881 & 8.815 & 5.635 & 3.532 \\
				$E_{\mathrm{5c}}$\hfill(eV) & 11.220 & 4.321 & 3.868 & 3.603 \\
				$E_{\mathrm{1u}}$\hfill(eV) & 27.510 & 12.640 & 10.369 & 9.009 \\
				$E_{\mathrm{3t}}$\hfill(eV) & 30.500 & 13.451 & 13.772 & 11.793 \\
				$E_{\mathrm{5d}}$\hfill(eV) & 31.067 & 17.082 & 15.290 & 12.239 \\
				$E_{\mathrm{1q}}$\hfill(eV) & 38.314 & 22.766 & 19.791 & 16.132 \\
				$\Delta_{\mathrm{v}}$\hfill(eV) & 0.024 & 0.046 & 0.226 & 0.362 \\
				$\Delta_{\mathrm{c}}$\hfill(eV) & 0.009 & 0.048 & 0.214 & 0.559 \\
				$\Delta_{\mathrm{d}}$\hfill(eV) & 0.003 & 0.004 & 0.006 & -0.004 \\
				$\Delta^-$  \hfill(eV) & -0.076i & 0.058i & -0.204i & 0.428i \\
				\hline
				$P_0$\hfill  (eV \AA) & 6.873 & 9.307 & 9.336 & 8.541 \\
				$P_1$\hfill  (eV \AA) & -3.211 & -1.730 & -2.074 & -1.942 \\
				$P_2$\hfill  (eV \AA) & 8.888 & 1.311 & 2.840 & 4.732 \\
				$P_3$\hfill  (eV \AA) & -6.343 & 12.586 & 11.276 & 9.429 \\
				$P_4$\hfill  (eV \AA) & 11.513 & 9.663 & 9.006 & 8.818 \\
				$P_5$\hfill  (eV \AA) & 2.846 & 2.085 & 2.245 & 2.055 \\
				$Q_0$\hfill  (eV \AA) & 11.277 & 9.128 & 8.892 & 8.714 \\
				$Q_1$\hfill  (eV \AA) & -8.968 & -6.551 & -6.481 & -5.649 \\
				$R_0$\hfill  (eV \AA) & 7.209 & 5.044 & 5.229 & 4.886 \\
				$R_1$\hfill  (eV \AA) & 10.538 & 7.333 & 6.850 & 6.043 \\
				$P'_0$\hfill (eV \AA) & 3.226i & -0.081i & 0.219i & -0.027i\\
				$P'_1$\hfill (eV \AA) & 2.920i & -1.277i & -1.508i & -0.681i \\
			\end{tabular}
		\end{ruledtabular}
	\end{table}
	
	\begin{table}
		\caption{\label{tab:AlX}Parameters for Al$X$ materials}
		\begin{ruledtabular}
			\begin{tabular}{lLLLL}
				& \multicolumn{1}{c}{AlN} & \multicolumn{1}{c}{AlP} & \multicolumn{1}{c}{AlAs} & \multicolumn{1}{c}{AlSb} \\
				\hline
				$E_{\mathrm{g}}$\hfill(eV) & 5.257 & 2.534 & 2.251 & 1.634 \\
				\hline
				$E_{\mathrm{1w}}$\hfill(eV) & -16.463 & -12.827 & -13.218 & -11.971 \\
				$E_{\mathrm{5v}}$\hfill(eV) & -0.007 & -0.022 & -0.108 & -0.218 \\
				$E_{\mathrm{1c}}$\hfill(eV) & 6.167 & 4.406 & 2.982 & 2.177 \\
				$E_{\mathrm{5c}}$\hfill(eV) & 14.906 & 5.748 & 5.297 & 3.717 \\
				$E_{\mathrm{1u}}$\hfill(eV) & 23.482 & 11.598 & 10.012 & 7.243 \\
				$E_{\mathrm{3t}}$\hfill(eV) & 20.288 & 10.299 & 10.258 & 8.514 \\
				$E_{\mathrm{5d}}$\hfill(eV) & 21.650 & 13.704 & 12.682 & 10.303 \\
				$E_{\mathrm{1q}}$\hfill(eV) & 28.545 & 17.451 & 15.954 & 13.266 \\
				$\Delta_{\mathrm{v}}$\hfill(eV) & 0.022 & 0.066 & 0.323 & 0.653 \\
				$\Delta_{\mathrm{c}}$\hfill(eV) & 0.051 & 0.029 & 0.038 & 0.060 \\
				$\Delta_{\mathrm{d}}$\hfill(eV) & 0.007 & 0.011 & 0.032 & 0.038 \\
				$\Delta^-$  \hfill(eV) & 0.002i & -0.015i & -0.116i & -0.257i \\
				\hline
				$P_0$\hfill  (eV \AA) & 8.231 & 8.571 & 8.871 & 8.750 \\
				$P_1$\hfill  (eV \AA) & -1.866 & -0.285 & 0.224 & 0.237 \\
				$P_2$\hfill  (eV \AA) & 2.632 & 2.512 & 2.336 & 1.781 \\
				$P_3$\hfill  (eV \AA) & 10.376 & 8.338 & 7.598 & 8.336 \\
				$P_4$\hfill  (eV \AA) & 11.197 & 9.346 & 8.805 & 7.602 \\
				$P_5$\hfill (eV \AA) & 1.465 & 1.884 & 1.563 & 2.108 \\
				$Q_0$\hfill  (eV \AA) & 10.036 & 8.078 & 8.068 & 7.518 \\
				$Q_1$\hfill  (eV \AA) & -7.351 & -4.649 & -4.372 & -4.338 \\
				$R_0$\hfill  (eV \AA) & 5.355 & 3.998 & 4.037 & 3.898 \\
				$R_1$\hfill  (eV \AA) & 7.820 & 6.395 & 6.331 & 5.637\\
				$P'_0$\hfill (eV \AA) & -2.391i & 0.632i & -0.579i & 0.580i \\
				$P'_1$\hfill (eV \AA) & -4.371i & 2.812i & -3.255i & -2.282i \\
			\end{tabular}
		\end{ruledtabular}
	\end{table}
	
	\begin{table}
		\label{tab:GaX}
		\caption{\label{tab:GaX}Parameters for Ga$X$ materials}
		\begin{ruledtabular}
			\begin{tabular}{lLLLL}
				& \multicolumn{1}{c}{GaN} & \multicolumn{1}{c}{GaP} & \multicolumn{1}{c}{GaAs} & \multicolumn{1}{c}{GaSb} \\
				\hline
				$E_{\mathrm{g}}$\hfill(eV) & 3.297 & 2.265 & 1.514 & 0.814 \\
				\hline
				$E_{\mathrm{1w}}$\hfill(eV) & -17.468 & -13.880 & -14.149 & -12.919 \\
				$E_{\mathrm{5v}}$\hfill(eV) & -0.011 & -0.033 & -0.126 & -0.244 \\
				$E_{\mathrm{1c}}$\hfill(eV) & 3.297 & 2.907 & 1.514 & 0.812 \\
				$E_{\mathrm{5c}}$\hfill(eV) & 12.289 & 4.840 & 4.754 & 3.496 \\
				$E_{\mathrm{1u}}$\hfill(eV) & 19.793 & 9.987 & 8.811 & 6.715 \\
				$E_{\mathrm{3t}}$\hfill(eV) & 21.157 & 10.945 & 11.267 & 9.478 \\
				$E_{\mathrm{5d}}$\hfill(eV) & 20.608 & 13.627 & 12.800 & 10.496 \\
				$E_{\mathrm{1q}}$\hfill(eV) & 24.682 & 16.900 & 15.662 & 13.193 \\
				$\Delta_{\mathrm{v}}$\hfill(eV) & 0.033 & 0.100 & 0.378 & 0.731 \\
				$\Delta_{\mathrm{c}}$\hfill(eV) & 0.315 & 0.169 & 0.191 & 0.219 \\
				$\Delta_{\mathrm{d}}$\hfill(eV) & -0.005 & 0.026 & 0.030 & 0.027 \\
				$\Delta^-$  \hfill(eV) & 0.029i & 0.041i & -0.038i & -0.217i \\
				\hline
				$P_0$\hfill  (eV \AA) & 7.511 & 8.904 & 9.343 & 9.298 \\
				$P_1$\hfill  (eV \AA) & -1.735 & -0.387 & 0.256 & 0.842 \\
				$P_2$\hfill  (eV \AA) & 5.344 & 2.511 & 2.152 & 1.421 \\
				$P_3$\hfill  (eV \AA) & 12.598 & 9.760 & 9.332 & 9.135 \\
				$P_4$\hfill  (eV \AA) & 11.963 & 8.863 & 8.372 & 7.534 \\
				$P_5$\hfill (eV \AA) & 3.067 & 2.499 & 2.389 & 2.379 \\
				$Q_0$\hfill  (eV \AA) & 10.308 & 8.228 & 8.350 & 7.981 \\
				$Q_1$\hfill  (eV \AA) & -6.405 & -5.464 & -5.106 & -4.424 \\
				$R_0$\hfill  (eV \AA) & 5.878 & 4.451 & 4.538 & 4.283 \\
				$R_1$\hfill  (eV \AA) & 5.152 & 6.154 & 6.095 & 5.691 \\
				$P'_0$\hfill (eV \AA) & 2.597i & 0.460i & -0.509i & 0.795i \\
				$P'_1$\hfill (eV \AA) & -3.407i & -1.955i & 2.455i & -1.576i \\
			\end{tabular}
		\end{ruledtabular}
	\end{table}
	
	\begin{table}
		\caption{\label{tab:InX}Parameters for In$X$ materials}
		\begin{ruledtabular}
			\begin{tabular}{lLLLL}
				& \multicolumn{1}{c}{InN} & \multicolumn{1}{c}{InP} & \multicolumn{1}{c}{InAs} & \multicolumn{1}{c}{InSb} \\
				\hline
				$E_{\mathrm{g}}$\hfill(eV) & 0.609 & 1.423 & 0.415 & 0.235 \\
				\hline
				$E_{\mathrm{1w}}$\hfill(eV) & -16.104 & -12.686 & -13.086 & -11.908 \\
				$E_{\mathrm{5v}}$\hfill(eV) & -0.014 & -0.041 & -0.134 & -0.254 \\
				$E_{\mathrm{1c}}$\hfill(eV) & 0.609 & 1.423 & 0.415 & 0.235 \\
				$E_{\mathrm{5c}}$\hfill(eV) & 10.954 & 4.889 & 4.710 & 3.500 \\
				$E_{\mathrm{1u}}$\hfill(eV) & 16.908 & 9.529 & 8.455 & 6.433 \\
				$E_{\mathrm{3t}}$\hfill(eV) & 17.700 & 10.253 & 10.392 & 8.832 \\
				$E_{\mathrm{5d}}$\hfill(eV) & 16.580 & 12.038 & 11.360 & 9.569 \\
				$E_{\mathrm{1q}}$\hfill(eV) & 19.848 & 14.717 & 13.783 & 11.964 \\
				$\Delta_{\mathrm{v}}$\hfill(eV) & 0.042 & 0.124 & 0.402 & 0.762 \\
				$\Delta_{\mathrm{c}}$\hfill(eV) & 0.721 & 0.435 & 0.447 & 0.411 \\
				$\Delta_{\mathrm{d}}$\hfill(eV) & -0.064 & 0.031 & 0.032 & 0.035 \\
				$\Delta^-$  \hfill(eV) & 0.055i & 0.139i & 0.001i & -0.061i \\
				\hline
				$P_0$\hfill  (eV \AA) & 6.636 & 7.913 & 8.394 & 8.553 \\
				$P_1$\hfill  (eV \AA) & -1.559 & -0.049 & 0.526 & 0.774 \\
				$P_2$\hfill  (eV \AA) & 4.267 & 3.215 & 2.768 & 1.846 \\
				$P_3$\hfill  (eV \AA) & 11.530 & 8.295 & 7.823 & 8.593 \\
				$P_4$\hfill  (eV \AA) & 10.839 & 8.610 & 8.166 & 7.164 \\
				$P_5$\hfill (eV \AA) & 3.706 & 2.163 & 1.777 & 2.105 \\
				$Q_0$\hfill  (eV \AA) & 9.519 & 7.905 & 7.987 & 7.573 \\
				$Q_1$\hfill  (eV \AA) & -6.803 & -5.036 & -4.632 & -4.294 \\
				$R_0$\hfill  (eV \AA) & 5.555 & 4.305 & 4.338 & 4.138 \\
				$R_1$\hfill  (eV \AA) & 3.813 & 5.618 & 5.675 & 5.268 \\
				$P'_0$\hfill (eV \AA) & 2.140i & -0.187i & -0.130i & -0.414i \\
				$P'_1$\hfill (eV \AA) & -3.648i & 2.609i & 3.188i & 2.256i \\
			\end{tabular}
		\end{ruledtabular}
	\end{table}

	In Figs.~\ref{fig:all_GaAs}(a,b), the $X$ and $Y$ element dependent values of the energy parameters can be seen. Since the $E_{8\mathrm{v}}$ is set to zero as the reference energy, the parameter $E_{5\mathrm{v}}$ is fixed to $-\Delta_\mathrm{v}/3$, which, in the energy scale of this plot, varies weakly with the atomic number.
	The behavior of the other parameters
	is rather systematic. Most of the conduction band energies decrease with the atomic number (of both $X$ and $Y$ element). However, there are some exceptions, e.g.
	the $E_{5\mathrm{c}}$ level in the $Y$As dependence, where it is almost constant or slightly increasing.
	The lowest valence band level $E_{1\mathrm{w}}$ increases, except to a slight decrease when changing from GaP to GaAs [Fig.~\ref{fig:all_GaAs}(a)] and from AlAs to GaAs [Fig.~\ref{fig:all_GaAs}(b)].
	
	In Figs.\ref{fig:all_GaAs}(c,d), the spin-orbit interaction parameters are shown. According to the adopted phase convention the $\Delta^-$ is purely imaginary. For the $X$ element dependence [Fig.\ref{fig:all_GaAs}(c)], the values of $\Delta_\mathrm{c}$, $\Delta_\mathrm{d}$ do not change much (except the difference between GaN and GaP for $\Delta_\mathrm{c}$). 
	The strongest dependence is for the valence band parameter $\Delta_\mathrm{v}$. It increases with the atomic number of $X$ and achieves the highest value of $0.731$~eV for GaSb.
	 These values ($\Delta_\mathrm{v}$) are in a good agreement with early predictions of Ref.~\cite{Herman1963}. Although, in the hydrogenic approximation, the spin-orbit interaction scales with the atomic number as $\propto Z^4$, it is known that this dependence is strongly suppressed by a screening~\cite{Shanavas2014}. In the case of $Y$As [Fig.\ref{fig:all_GaAs}(d)], one can observe a spectacularly increasing value of the $\Delta_\mathrm{c}$ parameter (except for the first two compounds), which in turn can be attributed to the increasing spin-orbit interaction in the $Y$ element, whose orbitals contribute more to the conduction band. On the other hand, the increase of $\Delta_\mathrm{v}$ is much weaker (as compared to the Ga$X$ dependence).
	 The increase of the absolute value of $\Delta^-$ from GaP to GaSb [Fig.\ref{fig:all_GaAs}(c)] and its decrease from AlAs to InAs [Fig.\ref{fig:all_GaAs}(d)] is consistent with the trends of the results in~Ref.~\cite{Jancu2005}.
	
	Finally, in Fig.~\ref{fig:all_GaAs}(e,f), one can see the results for the parameters related to momentum matrix elements. An interesting observation is that their values, except for few cases, vary rather weakly. In this case, however, it is rather difficult to find a universal key explaining this behavior.

	\begin{table}
		\caption{\label{tab:FitVsDFT}Comparison of the parameters obtained using  the fitting procedure and extracted directly from the DFT calculations, on the example of GaAs. We compare the moduli of the parameters related to the inversion asymmetry.}
		\begin{ruledtabular}
			\begin{tabular}{lLL}
				& \multicolumn{1}{c}{Fit}	&  \multicolumn{1}{c}{DFT} \\
				\hline
				${P_0}$\hfill  (eV \AA) & 9.343 & 9.704  \\
				${P_1}$\hfill  (eV \AA) & 0.256 & 0.050  \\
				${P_2}$\hfill  (eV \AA) & 2.152 & 1.579 \\
				${P_3}$\hfill  (eV \AA) & 9.332 & 9.337 \\
				${P_4}$\hfill  (eV \AA) & 8.372 & 8.414 \\
				${P_5}$\hfill  (eV \AA) & 2.389 & 2.344 \\
				${Q_0}$\hfill  (eV \AA) & 8.350 & 8.804 \\
				${Q_1}$\hfill  (eV \AA) & -5.106 & -5.030 \\
				${R_0}$\hfill  (eV \AA) & 4.538 & 4.634 \\
				${R_1}$\hfill  (eV \AA) & 6.095 & 5.651 \\
				$\abs{P'_0}$\hfill (eV \AA) & 0.509 & 1.668 \\
				$\abs{P'_1}$\hfill (eV \AA) & 2.455 & 3.079 \\
			\end{tabular}
		\end{ruledtabular}
	\end{table}
	
	In Table.~\ref{tab:FitVsDFT}, we compare the values of $P_{0...5}$, $P'_{0,1}$, $Q_{0,1}$, $R_{0,1}$ parameters for GaAs, calculated from the fitting and extracted directly from the DFT (as described in Sec.~\ref{sec:extr} and in Appendix~\ref{app:mel}). 
	We neglect the spin-orbit coupling in the determination of the DFT momentum matrix elements. Most of the fitted parameters are in a good agreement with the direct DFT approach. This shows that the effect of remote (not included in our $\kp$ model) bands is relatively weak. The differences get larger only for smaller parameters, like $P_1$ and $P_0'$. 
	
	There are also some, although limited, reference data in the literature. A comparison of the parameters evaluated in this work and those found in the literature (Refs.~\cite{Richard2004, Radhia_2007, doi:10.1063/1.3600643, doi:10.1063/1.2773532}) for GaAs, InP, InAs, InSb, AlAs and GaP is made in Tab.\ref{tab:Comparison}. One should note that different references could adopt various phase conventions. Therefore, to facilitate a comparison, we use the absolute values of $\Delta^-$ and also $E_{M} = \frac{2m_0}{\hbar^2}\abs{M}^2$, where $M$ are the parameters related to interband momentum matrix elements. The energy parameters in Tab.\ref{tab:Comparison} are given in the double-group notation, where
	\begin{align*}
		E_{6\alpha} &= E_{1\alpha}, \\
		E_{8\alpha} &= E_{5\alpha} + \frac{1}{3} \Delta_\alpha, \\
		E_{7\alpha} &= E_{5\alpha} - \frac{2}{3} \Delta_\alpha,
	\end{align*}
	with except to $\alpha = $``t'' where $E_{8\mathrm{t}} = E_{3\mathrm{t}}$.
	In most cases, the discrepancies of our results and the literature data are reasonable.
	Here however, we offer the parameters for all the $16$ compounds, obtained in a consistent way.
	
	For many III-V compounds with a direct band gap, the $8$-band $\bm{k}\cdot\bm{p}$ model is sufficient because the description of the electronic band structure around the $\Gamma$ point of the BZ is good enough. Therefore, the $30$-band $\bm{k}\cdot\bm{p}$ model is rarely used for GaAs, GaSb, InP, InAs and InSb, but it is already necessary for III-V compounds with an indirect band gap (B$X$, Al$X$, and GaP) in order to describe the side valleys correctly. The B$X$ compounds are experimentally the least studied ones, and the $\kp$ parameters for these compounds have not been reported so far.
	
	The electronic band structures for B$X$ compounds were calculated by DFT methods and reported in many articles \cite{Bouhafs_2000, FERHAT1998229, Zaoui_2000, CUI20091386, PhysRevB.98.081203}, but experimental studies of the band gap for cubic B$X$ are limited to a few cases   \cite{PhysRevLett.4.282, PhysRevLett.12.538, PhysRevB.101.035302, DAS2015439, DALUI2007149, YUE2020100194, Chu_1974}, mostly for BAs. The interest in BAs increased suddenly in 2013 when DFT calculations predicted BAs as a highly thermal conductive material with a thermal conductivity comparable with that of diamond \cite{PhysRevB.88.214303, PhysRevLett.111.025901}. Subsequent intense research led to a successful synthesis and verification of the predicted high thermal conductive material in 2018 by three groups \cite{Tian582, Kang575, Li579}, opening up new opportunities for both basic research and potential applications. A very important application of B$X$ compounds can be their alloying with the GaAs, InAs, and other III-V compounds. This aspect is interesting as it allows a wider range of strain engineering for some III-V alloys, especially GaAs-based semiconductors \cite{Kudrawiec2020}. 
	
	
	\renewcommand{\arraystretch}{1.2}
	\begin{table*}[t]
		\scriptsize
		\caption{\label{tab:Comparison}Comparison of our parameters and those presented in literature \cite{Richard2004, Radhia_2007, doi:10.1063/1.3600643, doi:10.1063/1.2773532}. For this, moduli of $\Delta^-$ and energies related to matrix elements $M$ by $E_{M} = \frac{2m_0}{\hbar^2}|M|^2$ were used. All parameters are in eV.}
		
		\begin{ruledtabular}
			\begin{tabular}{lLLLLLLLLLLLLL}
				& \multicolumn{1}{c}{GaAs$^a$} & \multicolumn{1}{c}{GaAs$^b$} & \multicolumn{1}{c}{InP$^a$} & \multicolumn{1}{c}{InP$^c$} & \multicolumn{1}{c}{InAs$^a$} & \multicolumn{1}{c}{InAs$^c$} & \multicolumn{1}{c}{InSb$^a$} & \multicolumn{1}{c}{InSb$^c$} & \multicolumn{1}{c}{AlAs$^a$} & \multicolumn{1}{c}{AlAs$^d$} & \multicolumn{1}{c}{AlAs$^e$} & \multicolumn{1}{c}{GaP$^a$} & \multicolumn{1}{c}{GaP$^e$}\\
				\hline
				$E_{\mathrm{6w}}$ & -14.149 & -12.55 & -12.686 & -11.078 & -13.086 & -12.69 & -11.908 & -11.71 & -13.218 & - & -11.95 & -13.880 & -12.30\\
				$E_{\mathrm{7v}}$ & -0.378 & -0.341 & -0.124 & -0.108 & -0.402 & -0.43 & -0.762 & -0.82 & -0.323 & - & -0.30 & -0.100 & -0.08\\
				$E_{\mathrm{8v}}$ & 0 & 0 & 0 & 0 & 0 & 0 & 0 & 0 & 0 & - & 0 & 0 & 0\\
				$E_{\mathrm{6c}}$ & 1.514 & 1.519 & 1.423 & 1.424 & 0.415 & 0.37 & 0.235 & 0.25 & 2.982 & - & 3.13 & 2.907 & 2.895\\
				$E_{\mathrm{7c}}$ & 4.626 & 4.488 & 4.599 & 4.72 & 4.412 & 4.39 & 3.227 & 3.16 & 5.271 & - & 4.54 & 4.728 & 4.87\\
				$E_{\mathrm{8c}}$ & 4.818 & 4.569 & 5.034 & 4.794 & 4.859 & 4.63 & 3.637 & 3.59 & 5.309 & - & 4.69 & 4.897 & 4.87\\
				$E_{\mathrm{6u}}$ & 8.811 & 8.56 & 9.529 & 8.50 & 8.455 & 8.55 & 6.433 & 8.56 & 10.012 & - & 9.89 & 9.987 & 8.80\\
				$E_{\mathrm{8t}}$ & 11.267 & 10.17 & 10.253 & 9.50 & 10.392 & 9.88 & 8.832 & 8.88 & 10.258 & - & 10.50 & 10.945 & 9.80\\
				$E_{\mathrm{7d}}$ & 12.780 & 11.89 & 12.017 & 11.50 & 11.338 & 11.89 & 9.545 & 9.89 & 12.660 & - & 12.50 & 13.610 & 11.80\\
				$E_{\mathrm{8d}}$ & 12.810 & 11.89 & 12.049 & 11.50 & 11.370 & 11.89 & 9.581 & 9.89 & 12.692 & - & 12.50 & 13.636 & 11.80\\
				$E_{\mathrm{6q}}$ & 15.662 & 13.64 & 14.717 & 12.99 & 13.783 & 12.64 & 11.964 & 12.64 & 15.954 & - & 13.64 & 16.900 & 13.30\\
				\hline
				$E_{\mathrm{P_0}}$ & 22.911 & 22.37 & 16.435 & 18.012 & 18.493 & 19.04 & 19.200 & 24.50 & 20.655 & 19.14 & 21.10 & 20.809 & 21.00\\
				$E_{\mathrm{Q_0}}$ & 18.298 & 16.79 & 16.402 & 14.01 & 16.742 & 15.64 & 15.052 & 14.50 & 17.086 & 14.29 & 16.80 & 17.770 & 17.01  \\
				$E_{\mathrm{R_0}}$ & 5.404 & 4.916 & 4.864 & 4.01 & 4.939 & 3.89 & 4.494 & 3.77 & 4.277 & 3.99 & 2.99 & 5.200 & 3.41\\
				$E_{\mathrm{P_2}}$ & 1.215 & 6.28 & 2.713 & 6.20 & 2.011 & 1.00 & 0.894 & 0.17 & 1.432 & 0.032 & 0 & 1.654 & 6.20\\
				$E_{\mathrm{P_1}}$ & 0.017 & 0.01 & 0.001 & 0.10 & 0.073 & 0.10 & 0.157 & 0.03 & 0.013 & 0.01 & 0.1 & 0.039 & 0.1\\
				$E_{\mathrm{Q_1}}$ & 6.842 & 4.344 & 6.658 & 7.50 & 5.630 & 5.00 & 4.840 & 2.34 & 5.016 & 8.49 & 4.12 & 7.835 & 7.00\\
				$E_{\mathrm{R_1}}$ & 9.751 & 8.888 & 8.283 & 11.15 & 8.452 & 11.66 & 7.283 & 8.60 & 10.520 & 9.29 & 13.06 & 9.940 & 12.50\\
				$E_{\mathrm{P_3}}$ & 22.857 & 23.15 & 18.060 & 2.51 & 16.063 & 2.50 & 19.378 & 7.99 & 15.153 & 15.01 & 3.50 & 25.004 & 2.51\\
				$E_{\mathrm{P_4}}$ & 18.396 & 19.63 & 19.458 & 20.05 & 17.503 & 19.00 & 13.471 & 16.00 & 20.347 & 16.00 & 18.00 & 20.617 & 20.05\\
				$E_{\mathrm{P_5}}$ & 1.498 & 2.434 & 1.228 & 3.43 & 0.829 & 5.00 & 1.163 & 0.40 & 0.641 & 1.79 & 0.10 & 1.639 & 3.43\\
				$E_{\mathrm{P_0'}}$ & 0.068 & 0.0656 & 0.009 & 0.15 & 0.004 & 0.01 & 0.045 & 0.03 & 0.088 & 0.14 & 0.16 & 0.055 & 0.50\\
				$E_{\mathrm{P_1'}}$ & 1.583 & 0 & 1.787 & 0 & 2.667 & 0 & 1.336 & 0 & 2.781 & 0 & 0 & 1.003 & 0\\
				$|\Delta^-|$   & 0.038 & 0 & 0.139 & 0.11 & 0.001 & 0.18 & 0.061 & 0.26 & 0.116 & 0 & 0 & 0.041 & 0\\
			\end{tabular}
		\end{ruledtabular}
		\raggedright$^a$This work. \par
		\raggedright$^b$Reference \cite{Richard2004}. \par
		\raggedright$^c$Reference \cite{Radhia_2007}. \par
		\raggedright$^d$Reference \cite{doi:10.1063/1.3600643}. \par
		\raggedright$^e$Reference \cite{doi:10.1063/1.2773532}. \par
	\end{table*}
	\section{Examples of application}
	\label{sec:examples}
	\begin{figure}[tb]
		\begin{center}
			\includegraphics[width=.48\textwidth]{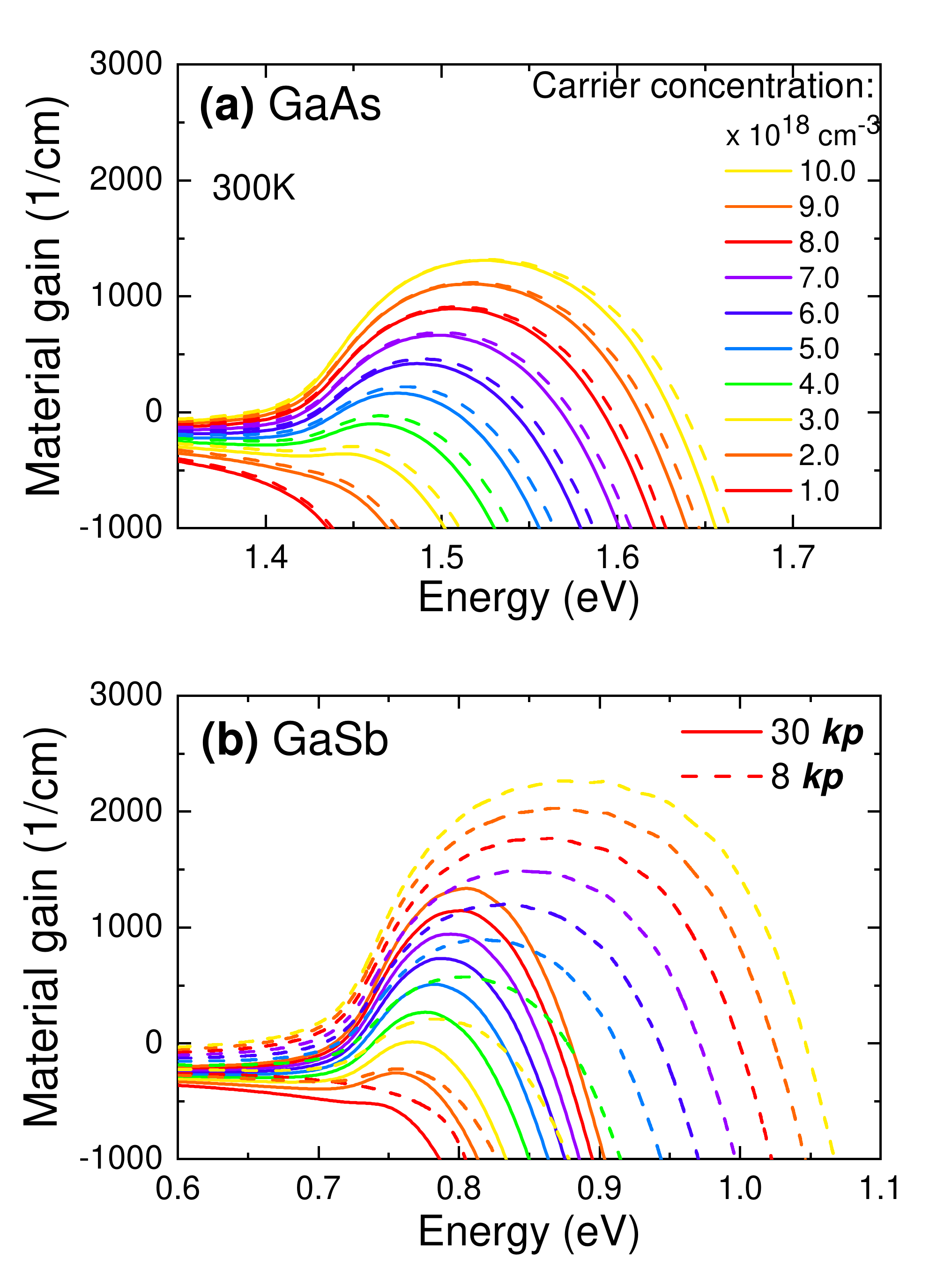}
		\end{center}
		\caption{\label{figgain} The material gain calculated for (a) GaAs and (b) GaSb under the $30$- and $8$-band $\kp$ model for carrier concentrations from: $1.0\cdot 10^{18}\mathrm{cm}^{-3}$ to $1.0\cdot 10^{19}\mathrm{cm}^{-3}$. The same colors correspond to a given carrier concentration. A solid line shows calculations of the $30$-band $\kp$ and a dashed line of the 8-band $\kp$ model. } 
	\end{figure}
	The optical gain for bulk materials was calculated as an exemplary application to highlight the importance of choice of the band model. The calculations were conducted for GaAs and GaSb using the $30$-band $\kp$ and the $8$-band $\kp$ models.
	Details regarding the calculation of the gain for these models are in the work \cite{pascha}. Here we calculate the $\kk$-dependent momentum matrix elements, using the Hellmann-Feynman theorem\cite{Feynman1939, LewYanVoon1993, eissfeller12}
	\begin{equation*}
		\label{eq:hf}
		\widetilde{P}_{n,ij}(\kk) \approx \frac{m_0}{\hbar} \pdv{H_{ij}(\kk)}{k_n},
	\end{equation*}
	where $H_{ij}$ are the matrix elements of the $8$- or $30$-band $\kp$ Hamiltonian.
	The parameters needed for calculations within the 8-band $\kp$ model: the effective mass and the Luttinger parameters were taken from the transformations presented in Appendix~\ref{app:reduction} (the reduced parameters denoted as $m'$ and $\gamma'_{1-3}$). Our objective was to perform calculations that will refer to the same band structure. Therefore, the calculations were performed with the parameters from Appendix~\ref{app:reduction}, despite the fact that for some materials they are inaccurate. The gain for both materials was calculated in the same range of carrier concentration from $1.0\cdot 10^{18} \mathrm{cm}^{-3}$ to $1.0\cdot 10^{19} \mathrm{cm}^{-3}$. The calculations were made for a temperature of $300$~K. The position of the conduction bands was shifted by $80$~meV in order to take the influence of temperature into account. The calculations made within the two models give very similar results for GaAs. We can notice a difference in the intensity of the calculated gain coefficient.  However, the shape of the spectrum and the maximum gain is almost the same. These differences become particularly noticeable for greater concentrations. This is due to the lack of consideration of the influence of the side valleys in the 8-band $\kp$ model. The situation is quite different for GaSb. The presented gain spectrum shows the legitimacy of using the larger-band model for gain calculations. The L valley is located very close to the $\Gamma$ valley in GaSb material. Therefore, the influence of the side valley is already visible for lower concentrations and practically should be taken into account in order to obtain correct results. It is also clear that, for semiconductors with an indirect energy gap, it is necessary to take into account the contribution of the side valleys for the calculation of the gain. This can be done by using the 30-band $\kp$ model because this model can satisfactory reflect the band structure of such materials.

	\section{Conclusions}
	\label{sec:concl}
	We have derived a ready-to-use symmetry invariant expansion of the $30$-band $\kp$ Hamiltonian. We have obtained its parameters for a wide class of III-V semiconductor compounds using the DFT band structure calculations as a reference. We have demonstrated a very good agreement between the DFT and the $\kp$ results. We have also discussed the chemical trends exhibited by the parameters. Finally, for GaAs, we have also compared the parameters obtained from the fitting and calculated directly from the DFT momentum matrix elements. We have shown a very good agreement for most of these values.
	\acknowledgments
	
	This work has been partially funded by a grant of
	the National Science Center Poland (OPUS11, UMO-
	2016/21/B/ST7/01267). Calculations have been carried
	out using resources provided by the Wroclaw Centre for Networking
	and Supercomputing and Interdisciplinary Centre for Mathematical and
	Computational Modeling (ICM) at the University of Warsaw.
	
	\appendix
	\section{Supplemental description and a derivation of the 30-band \texorpdfstring{$\bm{k}\!\vdot\!\bm{p}$}{p} Hamiltonian in the invariant expansion}
	\label{app:ham}
	In this Appendix, we describe the standard theory of invariants for the $\kp$ model~\cite{Bir1974,LewYanVoon2009}. Then, we use this approach to obtain an invariant expansion of the $30$-band Hamiltonian.
	
	A general form of the $\kp$ model is given by~\cite{LewYanVoon2009}
	\begin{equation}
		\label{eq:Hamiltonian_fundamental}
		H = H_0 + H_k + H_{\kp} + H_\mathrm{so},
	\end{equation}
	where
	\begingroup
	\allowdisplaybreaks
	\begin{align*}
		H_0  &= \frac{\hbar^2}{2m_0}p^2 + V(\rr),\\
		H_k  &= \frac{\hbar^2}{2m_0}k^2, \\
		H_{\kp} &= \frac{\hbar}{m_0} \bm{k} \cdot \bm{p} ,\\
		H_\mathrm{so} &= \frac{\hbar}{4 m^2_0 c^2 } (\nabla V \times \bm{p}) \cdot \bm{\sigma},
	\end{align*}
	\endgroup
	where $m_0$ is the free electron mass, $V(\rr)$ is a periodic crystal potential, $c$ is the speed of light, and $\bm{\sigma}$ is the vector of the Pauli matrices.
	Here, the spin-orbital terms proportional simultaneously to $\kk$ and $\nabla V$ are neglected.
	
	In a truncated (finite) basis, the Hamiltonian takes a matrix form. For the $30$-band model considered in this paper, we use the set of basis functions from Ref.~\onlinecite{Richard2004}, transformed to the form consistent with Ref.~\cite{Winkler2003}
	\begingroup
	\allowdisplaybreaks
	\begin{align*}
		\ket{\bm{\Gamma}_{6\mathrm{q}}} &=   \left\{
		\begin{array}{ll}
			\ket{S_\mathrm{q}} \otimes \ket{\uparrow}, \\[4pt]
			\ket{S_\mathrm{q}} \otimes \ket{\downarrow},
		\end{array} \right.\\[4pt]
		\ket{\bm{\Gamma}_{8\mathrm{d}}} &=   \left\{ \begin{array}{ll}
			-\frac{1}{\sqrt{2}} \ket{X_\mathrm{d}+iY_\mathrm{d}} \otimes \ket{\uparrow}, \\[4pt]
			\frac{2}{\sqrt{6}} \ket{Z_\mathrm{d}} \otimes \ket{\uparrow} - \frac{1}{\sqrt{6}} \ket{X_\mathrm{d}+iY_\mathrm{d}} \otimes \ket{\downarrow}, \\[4pt]
			\frac{1}{\sqrt{6}} \ket{X_\mathrm{d}-iY_\mathrm{d}} \otimes \ket{\uparrow} + \frac{2}{\sqrt{6}} \ket{Z_\mathrm{d}} \otimes \ket{\downarrow},  \\[4pt]
			\frac{1}{\sqrt{2}} \ket{X_\mathrm{d}-iY_\mathrm{d}} \otimes \ket{\downarrow},
		\end{array} \right.\\
		\ket{\bm{\Gamma}_{7\mathrm{d}}} &=   \left\{ \begin{array}{ll}
			-\frac{1}{\sqrt{3}} \ket{Z_\mathrm{d}} \otimes \ket{\uparrow} - \frac{1}{\sqrt{3}} \ket{X_\mathrm{d}+iY_\mathrm{d}} \otimes \ket{\downarrow}, \\[4pt]
			-\frac{1}{\sqrt{3}} \ket{X_\mathrm{d}-iY_\mathrm{d}} \otimes \ket{\uparrow} + \frac{1}{\sqrt{3}} \ket{Z_\mathrm{d}} \otimes \ket{\downarrow},  \\[4pt]
		\end{array} \right. \\
		\ket{\bm{\Gamma}_{8\mathrm{t}}} &=   \left\{ \begin{array}{ll}
			-\ket{D_{\mathrm{t},2}} \otimes \ket{\downarrow}, \\[4pt]
			\ket{D_{\mathrm{t},1}} \otimes \ket{\uparrow}, \\[4pt]
			-\ket{D_{\mathrm{t},1}} \otimes \ket{\downarrow}, \\[4pt]
			\ket{D_{\mathrm{t},2}} \otimes \ket{\uparrow}, \\[4pt]
		\end{array} \right. \\
		\ket{\bm{\Gamma}_{6\mathrm{u}}} &=   \left\{
		\begin{array}{ll}
			\ket{S_\mathrm{u}} \otimes \ket{\uparrow}, \\[4pt]
			\ket{S_\mathrm{u}} \otimes \ket{\downarrow},
		\end{array} \right.\\[4pt]
		\ket{\bm{\Gamma}_{8\mathrm{c}}} &=   \left\{ \begin{array}{ll}
			-\frac{1}{\sqrt{2}} \ket{X_\mathrm{c}+iY_\mathrm{c}} \otimes \ket{\uparrow}, \\[4pt]
			\frac{2}{\sqrt{6}} \ket{Z_\mathrm{c}} \otimes \ket{\uparrow} - \frac{1}{\sqrt{6}} \ket{X_\mathrm{c}+iY_\mathrm{c}} \otimes \ket{\downarrow}, \\[4pt]
			\frac{1}{\sqrt{6}} \ket{X_\mathrm{c}-iY_\mathrm{c}} \otimes \ket{\uparrow} + \frac{2}{\sqrt{6}} \ket{Z_\mathrm{c}} \otimes \ket{\downarrow},  \\[4pt]
			\frac{1}{\sqrt{2}} \ket{X_\mathrm{c}-iY_\mathrm{c}} \otimes \ket{\downarrow},
		\end{array} \right.\\
		\ket{\bm{\Gamma}_{7\mathrm{c}}} &=   \left\{ \begin{array}{ll}
			-\frac{1}{\sqrt{3}} \ket{Z_\mathrm{c}} \otimes \ket{\uparrow} - \frac{1}{\sqrt{3}} \ket{X_\mathrm{c}+iY_\mathrm{c}} \otimes \ket{\downarrow}, \\[4pt]
			-\frac{1}{\sqrt{3}} \ket{X_\mathrm{c}-iY_\mathrm{c}} \otimes \ket{\uparrow} + \frac{1}{\sqrt{3}} \ket{Z_\mathrm{c}} \otimes \ket{\downarrow},  \\[4pt]
		\end{array} \right. \\
		\ket{\bm{\Gamma}_{6\mathrm{c}}} &=   \left\{
		\begin{array}{ll}
			\ket{S_\mathrm{c}} \otimes \ket{\uparrow}, \\[4pt]
			\ket{S_\mathrm{c}} \otimes \ket{\downarrow},
		\end{array} \right.\\[4pt]
		\ket{\bm{\Gamma}_{8\mathrm{v}}} &=   \left\{ \begin{array}{ll}
			-\frac{1}{\sqrt{2}} \ket{X_\mathrm{v}+iY_\mathrm{v}} \otimes \ket{\uparrow}, \\[4pt]
			\frac{2}{\sqrt{6}} \ket{Z_\mathrm{v}} \otimes \ket{\uparrow} - \frac{1}{\sqrt{6}} \ket{X_\mathrm{v}+iY_\mathrm{v}} \otimes \ket{\downarrow}, \\[4pt]
			\frac{1}{\sqrt{6}} \ket{X_\mathrm{v}-iY_\mathrm{v}} \otimes \ket{\uparrow} + \frac{2}{\sqrt{6}} \ket{Z_\mathrm{v}} \otimes \ket{\downarrow},  \\[4pt]
			\frac{1}{\sqrt{2}} \ket{X_\mathrm{v}-iY_\mathrm{v}} \otimes \ket{\downarrow},
		\end{array} \right.\\
		\ket{\bm{\Gamma}_{7\mathrm{v}}} &=   \left\{ \begin{array}{ll}
			-\frac{1}{\sqrt{3}} \ket{Z_\mathrm{v}} \otimes \ket{\uparrow} - \frac{1}{\sqrt{3}} \ket{X_\mathrm{v}+iY_\mathrm{v}} \otimes \ket{\downarrow}, \\[4pt]
			-\frac{1}{\sqrt{3}} \ket{X_\mathrm{v}-iY_\mathrm{v}} \otimes \ket{\uparrow} + \frac{1}{\sqrt{3}} \ket{Z_\mathrm{v}} \otimes \ket{\downarrow},  \\[4pt]
		\end{array} \right. \\
		\ket{\bm{\Gamma}_{6\mathrm{w}}} &=   \left\{
		\begin{array}{ll}
			\ket{S_\mathrm{w}} \otimes \ket{\uparrow}, \\[4pt]
			\ket{S_\mathrm{w}} \otimes \ket{\downarrow},
		\end{array} \right.\\[4pt]
	\end{align*}
	\endgroup
	where
	$\ket{S}$, $\ket{X}$, $\ket{Y}$, $\ket{Z}$, $\ket{D_1} = \ket{2Z^2-X^2-Y^2}$, $\ket{D_{2}} =  \ket{\sqrt{3}\qty(X^2-Y^2)}$ describe the orbital symmetry, while $\ket{\uparrow/\downarrow}$ denotes the spin projection. One should note, that the states $\{ \ket{X}, \ket{Y}, \ket{Z} \}$ transform like $\{x,y,z\}$ and $\{yz,zx,xy\}$~\cite{LewYanVoon2009}. We assume that the wave functions for the ``w",``v",``u",``d" bands are real, while for the ``c",``t",``q" bands they are purely imaginary. In the applied phase convention, the parameters that appear due to inversion asymmetry are imaginary~\cite{Winkler2003}.

	To obtain the Hamiltonian in a block form, it is convenient to perform calculations in a spherical basis with Eq.~\ref{eq:Hamiltonian_fundamental} written in terms of spherical tensor components~\cite{LewYanVoon2009, Machnikowski2019}. After a simple algebra, one can obtain
	\begin{align*}
		H_0  &= \frac{\hbar^2}{2m_0}  \qty[ (p^{(1)}_{0} )^2 - 2 \, p^{(1)}_{-1} p^{(1)}_{1} ] + V(\rr),\\
		H_k  &= \frac{\hbar^2}{2m_0}  \qty[ (k^{(1)}_{0} )^2 - 2 \, k^{(1)}_{-1} k^{(1)}_{1} ], \\
		H_{\kp} & = \frac{\hbar}{m_0} \qty[ k^{(1)}_{0} p^{(1)}_{0}  - k^{(1)}_{-1} p^{(1)}_{1} - k^{(1)}_{1} p^{(1)}_{-1} ], \\
		H_{\mathrm{so}} & = \frac{\hbar}{4 m^2_0 c^2 } \qty[ X^{(1)}_{0} \sigma^{(1)}_{0}  - X^{(1)}_{-1} \sigma^{(1)}_{1} - X^{(1)}_{1} \sigma^{(1)}_{-1} ],
	\end{align*}
	where $v^{(1)}_{q}$ are the spherical vector components (with the upper index as the tensor rank) defined with respect to the cartesian ones in the standard way:
	\begin{align*}
		v^{(1)}_{-1} &= \frac{1}{\sqrt{2}} \qty(v_x - i v_y),\\
		v^{(1)}_{0}  &= v_z, \\
		v^{(1)}_{1}  &= -\frac{1}{\sqrt{2}} \qty(v_x + i v_y).
	\end{align*}
	The vector $X^{(1)}$ can be calculated using tensor multiplication rules giving
	$$
	X^{(1)}_{q} = -i \sqrt{2} \sum_{q_1,q_2} (\nabla V)^{(1)}_{q_1} p^{(1)}_{q_2} \braket{1 1; q_1 q_2}{1q},
	$$
	where $\braket{j_1 j_2; m_1 m_2}{J M}$ are the Clebsch-Gordan coefficients.
	Finally, $\sigma^{(1)}_q$ are components of the spherical vector built from the Pauli matrices (see the calculations in Appendix of Ref.~\cite{Machnikowski2019} as an illustration).
	The basis functions can be further expressed by the angular momentum eigenvectors $\ket{\widetilde{\alpha}; lm}$ (which refer to spherical harmonics $Y^m_{l}$)~\cite{LewYanVoon2009,Machnikowski2019}
	$$
	\ket{\Gamma_{\widetilde{\alpha},i}} = \sum_{l,m,s} c^{(\widetilde{\alpha},i)}_{lms} \ket{\widetilde{\alpha}; lm} \otimes \ket{s},
	$$
	where $\ket{\Gamma_{\widetilde{\alpha},i}}$ is the $i$-th component of the pseudospinor $\ket{\bm{\Gamma}_{\widetilde{\alpha}}}$, the $c^{(\widetilde{\alpha},i)}_{lms}$ are coefficients, and the common index $\widetilde{\alpha} \equiv \{n,\alpha\}$ denotes the representation and the band set.
	Finally, with the Wigner-Eckart theorem~\cite{Trebin1979,LewYanVoon2009,Machnikowski2019}, the Hamiltonian blocks $H_{\widetilde{\alpha},\widetilde{\beta}} = \mel{\bm{\Gamma}_{\widetilde{\alpha}}}{H}{\bm{\Gamma}_{\widetilde{\beta}}}$ are calculated, where $H_0$ and $H_k$ give rise only to the diagonal parts of $H_{\widetilde{\alpha},\widetilde{\alpha}}$. The full $30$-band Hamiltonian can be presented in a block matrix form
	\begin{widetext}
		\begin{equation}
			\label{eq:Hblock}
			H=
			\left(
			\begin{array}{*{11}{c}}
				H_{\mathrm{6q6q}} & H_{\mathrm{6q8d}} & H_{\mathrm{6q7d}} & H_{\mathrm{6q8t}} & H_{\mathrm{6q6u}} & H_{\mathrm{6q8c}} & H_{\mathrm{6q7c}} & H_{\mathrm{6q6c}} & H_{\mathrm{6q8v}} & H_{\mathrm{6q7v}} & H_{\mathrm{6q6w}} \\
				H_{\mathrm{8d6q}} & H_{\mathrm{8d8d}} & H_{\mathrm{8d7d}} & H_{\mathrm{8d8t}} & H_{\mathrm{8d6u}} & H_{\mathrm{8d8c}} & H_{\mathrm{8d7c}} & H_{\mathrm{8d6c}} & H_{\mathrm{8d8v}} & H_{\mathrm{8d7v}} & H_{\mathrm{8d6w}} \\
				H_{\mathrm{7d6q}} & H_{\mathrm{7d8d}} & H_{\mathrm{7d7d}} & H_{\mathrm{7d8t}} & H_{\mathrm{7d6u}} & H_{\mathrm{7d8c}} & H_{\mathrm{7d7c}} & H_{\mathrm{7d6c}} & H_{\mathrm{7d8v}} & H_{\mathrm{7d7v}} & H_{\mathrm{7d6w}} \\
				H_{\mathrm{8t6q}} & H_{\mathrm{8t8d}} & H_{\mathrm{8t7d}} & H_{\mathrm{8t8t}} & H_{\mathrm{8t6u}} & H_{\mathrm{8t8c}} & H_{\mathrm{8t7c}} & H_{\mathrm{8t6c}} & H_{\mathrm{8t8v}} & H_{\mathrm{8t7v}} & H_{\mathrm{8t6w}} \\
				H_{\mathrm{6u6q}} & H_{\mathrm{6u8d}} & H_{\mathrm{6u7d}} & H_{\mathrm{6u8t}} & H_{\mathrm{6u6u}} & H_{\mathrm{6u8c}} & H_{\mathrm{6u7c}} & H_{\mathrm{6u6c}} & H_{\mathrm{6u8v}} & H_{\mathrm{6u7v}} & H_{\mathrm{6u6w}} \\
				H_{\mathrm{8c6q}} & H_{\mathrm{8c8d}} & H_{\mathrm{8c7d}} & H_{\mathrm{8c8t}} & H_{\mathrm{8c6u}} & \tikzmarkin[disable rounded corners=true,color=lightgray]{14kp}(0.05,-0.09)(-0.15,0.31)  H_{\mathrm{8c8c}} & H_{\mathrm{8c7c}} & H_{\mathrm{8c6c}} & H_{\mathrm{8c8v}} & H_{\mathrm{8c7v}} & H_{\mathrm{8c6w}} \\
				H_{\mathrm{7c6q}} & H_{\mathrm{7c8d}} & H_{\mathrm{7c7d}} & H_{\mathrm{7c8t}} & H_{\mathrm{7c6u}} & H_{\mathrm{7c8c}} & H_{\mathrm{7c7c}} & H_{\mathrm{7c6c}} & H_{\mathrm{7c8v}} & H_{\mathrm{7c7v}} & H_{\mathrm{7c6w}} \\
				H_{\mathrm{6c6q}} & H_{\mathrm{6c8d}} & H_{\mathrm{6c7d}} & H_{\mathrm{6c8t}} & H_{\mathrm{6c6u}} & H_{\mathrm{6c8c}} & H_{\mathrm{6c7c}} & \tikzmarkin[disable rounded corners=true,color=black,fill=lightgray]{8kp}(0.05,-0.09)(-0.15,0.31)  \tikzmarkin[disable rounded corners=true,color=newcyan]{2kp}(0.05,-0.09)(-0.15,0.31) H_{\mathrm{6c6c}} \tikzmarkend{2kp} & H_{\mathrm{6c8v}} & H_{\mathrm{6c7v}} & H_{\mathrm{6c6w}} \\
				H_{\mathrm{8v6q}} & H_{\mathrm{8v8d}} & H_{\mathrm{8v7d}} & H_{\mathrm{8v8t}} & H_{\mathrm{8v6u}} & H_{\mathrm{8v8c}} & H_{\mathrm{8v7c}} & H_{\mathrm{8v6c}} & \tikzmarkin[disable rounded corners=true,color=newred]{6kp}(0.05,-0.09)(-0.15,0.31) H_{\mathrm{8v8v}} & H_{\mathrm{8v7v}} & H_{\mathrm{8v6w}} \\
				H_{\mathrm{7v6q}} & H_{\mathrm{7v8d}} & H_{\mathrm{7v7d}} & H_{\mathrm{7v8t}} & H_{\mathrm{7v6u}} & H_{\mathrm{7v8c}} & H_{\mathrm{7v7c}} & H_{\mathrm{7v6c}} & H_{\mathrm{7v8v}} & H_{\mathrm{7v7v}} \tikzmarkend{6kp}   \tikzmarkend{8kp} \tikzmarkend{14kp}   & H_{\mathrm{7v6w}} \\
				H_{\mathrm{6w6q}} & H_{\mathrm{6w8d}} & H_{\mathrm{6w7d}} & H_{\mathrm{6w8t}} & H_{\mathrm{6w6u}} & H_{\mathrm{6w8c}} & H_{\mathrm{6w7c}} & H_{\mathrm{6w6c}} & H_{\mathrm{6w8v}} & H_{\mathrm{6w7v}} & H_{\mathrm{6w6w}}
			\end{array}
			\right),
		\end{equation}
	\end{widetext}
	where we show the order of bands relevant for most of the considered III-V semiconductors (like in GaAs). The block positions can be related to fewer-band models that are often used in the semiconductor physics. The part of Eq.$\ref{eq:Hblock}$ which is highlighted by the blue color ($H_{\mathrm{6c6c}}$) describes the lowest conduction band and corresponds to the $2$-band $\kp$ Hamiltonian~\cite{Mielnik-Pyszczorski2018}. The part marked by the red color contains the highest valence band block $H_{\mathrm{8v8v}}$ describing the heavy and light hole, and the spin-orbit split-off bands $H_{\mathrm{7v7v}}$. The part separated by the solid line corresponds to the standard $8$-band $\kp$ model~\cite{bahder90}, while the gray area extends the model to $14$ bands~\cite{Trebin1979, Winkler2003}. One should recall, that reduction of the full $30$-band $\kp$ model to a model with fewer number of bands involves a perturbative approach, as described in Appendix~\ref{app:reduction}.
	
	The invariant expansion of the 30-band $\kp$ Hamiltonian can be written in an explicit form
	\begingroup
	\allowdisplaybreaks
	\begin{align*}
		H_{\mathrm{6q6q}} &=  \qty( E_{\mathrm{1q}} + \frac{\hbar^2}{2m_0} k^2 ) \; \mathbb{I}_2 ,\\
		H_{\mathrm{8d8d}} &=  \qty( E_{\mathrm{5d}} + \frac{1}{3} \Delta_{\mathrm{d}} + \frac{\hbar^2}{2m_0} k^2 ) \; \mathbb{I}_4 ,\\
		H_{\mathrm{7d7d}} &=  \qty( E_{\mathrm{5d}} - \frac{2}{3} \Delta_{\mathrm{d}} + \frac{\hbar^2}{2m_0} k^2 ) \; \mathbb{I}_2 ,\\
		H_{\mathrm{8t8t}} &=  \qty( E_{\mathrm{3t}} + \frac{\hbar^2}{2m_0} k^2 )\; \mathbb{I}_4 ,\\
		H_{\mathrm{6u6u}} &=  \qty( E_{\mathrm{1u}} + \frac{\hbar^2}{2m_0} k^2 )\; \mathbb{I}_2 ,\\
		H_{\mathrm{8c8c}} &=  \qty( E_{\mathrm{5c}} + \frac{1}{3} \Delta_{\mathrm{c}} + \frac{\hbar^2}{2m_0} k^2 ) \; \mathbb{I}_4 ,\\
		H_{\mathrm{7c7c}} &=  \qty( E_{\mathrm{5c}} - \frac{2}{3} \Delta_{\mathrm{c}} + \frac{\hbar^2}{2m_0} k^2 ) \; \mathbb{I}_2 ,\\
		H_{\mathrm{6c6c}} &=  \qty( E_{\mathrm{1c}} + \frac{\hbar^2}{2m_0} k^2 ) \; \mathbb{I}_2 ,\\
		H_{\mathrm{8v8v}} &=  \qty( E_{\mathrm{5v}} + \frac{1}{3} \Delta_{\mathrm{v}} + \frac{\hbar^2}{2m_0} k^2 ) \; \mathbb{I}_4 ,\\
		H_{\mathrm{7v7v}} &=  \qty( E_{\mathrm{5v}} - \frac{2}{3} \Delta_{\mathrm{v}} + \frac{\hbar^2}{2m_0} k^2 ) \; \mathbb{I}_2 ,\\
		H_{\mathrm{6w6w}} &=  \qty( E_{\mathrm{1w}} + \frac{\hbar^2}{2m_0} k^2 ) \; \mathbb{I}_2 ,\\
		H_{\mathrm{6q8d}} &=  \sqrt{3} P_3 \qty(T_x k_x + \cp),\\
		H_{\mathrm{6q7d}} &= -\frac{1}{\sqrt{3}} P_3 \qty(\sigma_x k_x + \cp),\\
		H_{\mathrm{6q8v}} &= \sqrt{3} P_2 (T_x k_x + \cp),\\
		H_{\mathrm{6q7v}} &= -\frac{1}{\sqrt{3}} P_2 (\sigma_x k_x + \cp),\\
		H_{\mathrm{8d8t}} &= -\sqrt{30} \ R_1 (D^\dagger_x k_x + \cp),\\
		H_{\mathrm{8d8c}} &= -\frac{2}{3} Q_1 \qty(\{J_y,J_z\} k_x + \cp),\\
		H_{\mathrm{8d7c}} &= -2 Q_1 (T^\dagger_{yz} k_x+\cp),\\
		H_{\mathrm{8d6c}} &= \sqrt{3} P_1 (T^\dagger_x k_x+\cp),\\
		H_{\mathrm{7d8t}} &= \sqrt{6} R_1 \qty(T_x k_x+\cp),\\
		H_{\mathrm{7d8c}} &= -2 Q_1 (T_{yz} k_x+\cp),\\
		H_{\mathrm{7d6c}} &= -\frac{1}{\sqrt{3}} P_1 (\sigma_x k_x+\cp),\\
		H_{\mathrm{8t8v}} &= -\sqrt{30} R_0 \qty(D_x k_x+\cp),\\
		H_{\mathrm{8t7v}} &= \sqrt{6} R_0 \qty(T^\dagger_x k_x+\cp),\\
		H_{\mathrm{6u8c}} &= \sqrt{3} P_4 (T_x k_x+\cp),\\
		H_{\mathrm{6u7c}} &= -\frac{1}{\sqrt{3}} P_4 \qty(\sigma_x k_x+\cp),\\
		H_{\mathrm{8c6c}} &= \sqrt{3} P'^*_0 \qty(T^\dagger_x k_x+\cp),\\
		H_{\mathrm{8c8v}} &= -\frac{2}{3} Q_0 \qty(\{J_y,J_z\} k_x+\cp) + \frac{1}{3} \Delta^{-} \; \mathbb{I}_4 ,\\
		H_{\mathrm{8c7v}} &= -2 Q_0 \qty(T^\dagger_{yz} k_x+\cp),\\
		H_{\mathrm{8c6w}} &= \sqrt{3} P_5 \qty(T^\dagger_x k_x+\cp),\\
		H_{\mathrm{7c6c}} &= -\frac{1}{\sqrt{3}} P'^*_0 (\sigma_x k_x+\cp),\\
		H_{\mathrm{7c8v}} &= -2 Q_0 \qty(T_{yz} k_x+\cp),\\
		H_{\mathrm{7c7v}} &= -\frac{2}{3} \Delta^{-} \; \mathbb{I}_2, \\
		H_{\mathrm{7c6w}} &= -\frac{1}{\sqrt{3}} P_5 (\sigma_x k_x+\cp),\\
		H_{\mathrm{6c8v}} &= \sqrt{3} P_0 \qty(T_x k_x+\cp),\\
		H_{\mathrm{6c7v}} &= -\frac{1}{\sqrt{3}} P_0 (\sigma_x k_x+\cp),\\
		H_{\mathrm{8v6w}} &= \sqrt{3} P'^*_1 \qty(T^\dagger_x k_x+\cp),\\
		H_{\mathrm{7v6w}} &= -\frac{1}{\sqrt{3}} P'^*_1 (\sigma_x k_x+\cp),		
	\end{align*}
	\endgroup
	where the matrices $T_i$ and $J_i$ are~\cite{Winkler2003,Trebin1979}
	\begingroup
	\allowdisplaybreaks
	\begin{align*}
		T_x & = \frac{\sqrt{2}}{6}
		\begin{pmatrix}
			-\sqrt{3} & 0 & 1 & 0 \\
			0 & -1 & 0 & \sqrt{3}
		\end{pmatrix},\\
		T_y & = -\frac{i \sqrt{2}}{6}
		\begin{pmatrix}
			\sqrt{3} & 0 & 1 & 0 \\
			0 & 1 & 0 & \sqrt{3}
		\end{pmatrix},\\
		T_z & = \frac{\sqrt{2}}{3}
		\begin{pmatrix}
			0 & 1 & 0 & 0 \\
			0 & 0 & 1 & 0
		\end{pmatrix},\\
		J_x & = \frac{1}{2}
		\begin{pmatrix}
			0 & \sqrt{3} & 0 & 0 \\
			\sqrt{3} & 0 & 2 & 0 \\
			0 & 2 & 0 & \sqrt{3} \\
			0 & 0 & \sqrt{3} & 0
		\end{pmatrix},\\
		J_y & = \frac{i}{2}
		\begin{pmatrix}
			0 & -\sqrt{3} & 0 & 0 \\
			\sqrt{3} & 0 & -2 & 0 \\
			0 & 2 & 0 & -\sqrt{3} \\
			0 & 0 & \sqrt{3} & 0
		\end{pmatrix},\\
		J_z & = \frac{1}{2}
		\begin{pmatrix}
			3 & 0 & 0 & 0 \\
			0 & 1 & 0 & 0 \\
			0 & 0 & -1 & 0 \\
			0 & 0 & 0 & -3
		\end{pmatrix},
	\end{align*}
	\endgroup
	and the $T_{ij}$ matrices are defined as $T_{ij} = T_i J_j + T_j J_i$.
	Finally, we found the matrices $D_i$ from the Wigner-Eckart theorem, which give
	\begingroup
	\allowdisplaybreaks
	\begin{align*}
		D_x & = \frac{1}{6 \sqrt{5}}
		\begin{pmatrix}
			0 & \sqrt{3} & 0 & -3 \\
			\sqrt{3} & 0 & -1 & 0 \\
			0 & -1 & 0 & \sqrt{3} \\
			-3 & 0 & \sqrt{3} & 0
		\end{pmatrix},\\
		D_y & = \frac{i}{6 \sqrt{5}}
		\begin{pmatrix}
			0 & -\sqrt{3} & 0 & -3 \\
			\sqrt{3} & 0 & 1 & 0 \\
			0 & -1 & 0 & -\sqrt{3} \\
			3 & 0 & \sqrt{3} & 0
		\end{pmatrix},\\
		D_z & = \frac{1}{3 \sqrt{5}}
		\begin{pmatrix}
			0 & 0 & 0 & 0 \\
			0 & 2 & 0 & 0 \\
			0 & 0 & -2 & 0 \\
			0 & 0 & 0 & 0
		\end{pmatrix}.
	\end{align*}
	\endgroup

	\section{Reduction to fewer band models}
	\label{app:reduction}
	The $30$-band $\kp$ model can be further reduced to fewer band models, using the L\"owdin perturbation theory~\cite{Lowdin1950,LewYanVoon2009}.	
	Here, we show the standard reduction formulas within the parameter convention consistent to our paper. In further calculations, we utilize the following relations:
	\begin{align*}
		\{\sigma_i, \sigma_j\} &=  \mathbb{I}_2 \delta_{ij}, \\
		(T_i T^\dagger_j + T_j T^\dagger_i)/2 &= \frac{2}{9} \mathbb{I}_2 \delta_{ij}, \\
		(T^\dagger_i T_j + T^\dagger_j T_i)/2 &= \frac{1}{9} \mathbb{I}_4 \delta_{ij} - \frac{1}{9} \qty(  J^2_i - \frac{J^2}{3} ) \delta_{ij} \\ &- \frac{1}{9} \{J_i,J_j\} (1- \delta_{ij} ), \\
		\{J_i, J_j\} &= \frac{5}{4} \mathbb{I}_4  \delta_{ij} + \qty(  J^2_i - \frac{J^2}{3} ) \delta_{ij} \\ &+ \{J_i,J_j\} (1- \delta_{ij} ), \\
		\{D_i, D_j\} &= \frac{2}{45} \mathbb{I}_4  \delta_{ij} - \frac{2}{45} \qty(  J^2_i - \frac{J^2}{3} ) \delta_{ij} \\ & + \frac{1}{45} \{J_i,J_j\} (1- \delta_{ij} ). \\
	\end{align*}	
	The symmetric product is defined as $\{A,B\} = (AB+BA)/2$.
	We use also:
	\begin{align*}
		\{J_x,J_y\} \{J_x,J_y\}  &= \frac{3}{4}  \mathbb{I}_4, \\
		\{\{J_x,J_y\}, \{J_y,J_z\} \} &= 0, \\
		T^\dagger_{xy} T_{xy}  &= \frac{1}{12}  \mathbb{I}_4 + \frac{1}{12} \qty(  J^2_z - \frac{J^2}{3} ), \\
		T^\dagger_{yz} T_{zx} + T^\dagger_{zx} T_{yz} &= -\frac{1}{6} \{J_x,J_y\}, \\
	\end{align*}
	which holds under cyclic permutation of indices.
	
	To obtain the electron effective mass, one need to account the remote band contributions to the $H_\mathrm{6c6c}$ block coming from the coupling to the $\Gamma_8$ and $\Gamma_7$ bands (while the ones from $\Gamma_6$ vanish)~\cite{eissfeller12,Mielnik-Pyszczorski2018}
	\begin{align*}
		H^\mathrm{(r)}_\mathrm{6c6c} & =  \sum_{\alpha} \frac{H_{6c8\alpha} H_{8\alpha 6c}}{E_{6\mathrm{c}} - E_{8\mathrm{\alpha}}} + \sum_{\alpha} \frac{H_{6c7\alpha} H_{7\alpha 6c}}{E_{6\mathrm{c}} - E_{7\mathrm{\alpha}}}. 
	\end{align*}
	Since we do not consider the magnetic field (hence $[k_i,k_j] = 0$), only the symmetric products of the involved vector components give rise to the sum~\cite{Mielnik-Pyszczorski2018}
	\begin{align*}
		H^\mathrm{(r)}_\mathrm{6c6c} & = 3 \sum_{\alpha} \sum_{i,j} \frac{\abs{\mathcal{P}_{c\alpha}}^2}{E_{6\mathrm{c}} - E_{8\mathrm{\alpha}}}  \{k_i, k_j\}  \frac{T_i T^\dagger_j + T_j T^\dagger_i}{2} \\ & + \frac{1}{3} \sum_{\alpha} \sum_{i,j} \frac{\abs{\mathcal{P}_{c\alpha}}^2}{E_{6\mathrm{c}} - E_{7\mathrm{\alpha}}}  \{k_i, k_j\}  \{\sigma_i, \sigma_j\},
	\end{align*}
	giving
	\begin{align*}
		H^\mathrm{(r)}_\mathrm{6c6c} & = \frac{2}{3} \sum_{\alpha} \frac{\abs{\mathcal{P}_{c\alpha}}^2}{E_{6\mathrm{c}} - E_{8\mathrm{\alpha}}}  k^2 \mathbb{I}_2  + \frac{1}{3} \sum_{\alpha} \frac{\abs{\mathcal{P}_{c\alpha}}^2}{E_{6\mathrm{c}} - E_{7\mathrm{\alpha}}}  k^2 \mathbb{I}_2.
	\end{align*}
	This allows to extract the electron effective mass in a form
	$$
	\frac{m_0}{m^*} = 1 + \frac{2 m_0}{\hbar^2} \sum_{\alpha} \abs{\mathcal{P}_{c\alpha}}^2 \qty[ \frac{2}{3} \frac{1}{E_{6\mathrm{c}} - E_{8\mathrm{\alpha}}} + \frac{1}{3}  \frac{1}{E_{6\mathrm{c}} - E_{7\mathrm{\alpha}}}  ].
	$$
	Hence, in our model, one can obtain
	\begin{align}
		\frac{m_0}{m^*} &= 1 + \frac{2 m_0}{\hbar^2} \abs{P_{0}}^2 \qty[ \frac{2}{3} \frac{1}{E_{6\mathrm{c}} - E_{8\mathrm{v}}} + \frac{1}{3}  \frac{1}{E_{6\mathrm{c}} - E_{7\mathrm{v}}}  ] \nonumber
		\\ & + \frac{2 m_0}{\hbar^2} \abs{P_{1}}^2 \qty[ \frac{2}{3} \frac{1}{E_{6\mathrm{c}} - E_{8\mathrm{d}}} + \frac{1}{3}  \frac{1}{E_{6\mathrm{c}} - E_{7\mathrm{d}}}  ] \nonumber
		\\ & + \frac{2 m_0}{\hbar^2} \abs{P'_{0}}^2 \qty[ \frac{2}{3} \frac{1}{E_{6\mathrm{c}} - E_{8\mathrm{c}}} + \frac{1}{3}  \frac{1}{E_{6\mathrm{c}} - E_{7\mathrm{c}}}  ].
		\label{eq:paramEM}
	\end{align}

	We perform a procedure similar as described above to obtain the Luttinger parameters for the $H_\mathrm{8v8v}$ band block.	
	The terms are grouped and compared to the form~\cite{Winkler2003,eissfeller12}
	\begin{align*}
		H_\mathrm{8v8v} = -\frac{\hbar^2}{2m_0} \Bigg [ \gamma_1 k_x^2 - 2 \gamma_2 \qty( J^2_x -\frac{J^2}{3}) k^2_x \\ - 4 \gamma_3 \{J_x, J_y\} \{k_x, k_y\} + \mathrm{c.p.} \Bigg ].
	\end{align*}
	This leads to the formulas:
	\begin{align*}
		\gamma_1 =& -1 + \frac{2 m_0}{ \hbar^2} \sum_{\alpha  \neq \mathrm{v}} \Bigg [  \frac{1}{3} \frac{\abs{\mathcal{P}_{\mathrm{v} \alpha}}^2}{ E_{6\alpha} - E_{8\mathrm{v}}} +  \frac{1}{3} \frac{\abs{\mathcal{Q}_{\mathrm{v} \alpha}}^2}{  E_{7\alpha} - E_{8\mathrm{v}}}  \\
		& + \frac{1}{3} \frac{\abs{\mathcal{Q}_{\mathrm{v} \alpha}}^2}{E_{8\alpha} - E_{8\mathrm{v}}}
		+ \frac{4}{3} \frac{\abs{\mathcal{R}_{\mathrm{v} \alpha}}^2}{ E_{8\alpha} - E_{8\mathrm{v}}}
		\Bigg ],\\
		\gamma_2 =& \frac{2 m_0}{ \hbar^2} \sum_{\alpha  \neq \mathrm{v}} \Bigg [  \frac{1}{6} \frac{\abs{\mathcal{P}_{\mathrm{v} \alpha}}^2}{ E_{6\alpha} - E_{8\mathrm{v}}} -  \frac{1}{6} \frac{\abs{\mathcal{Q}_{\mathrm{v} \alpha}}^2}{  E_{7\alpha} - E_{8\mathrm{v}}}  \\
		& + \frac{2}{3} \frac{\abs{\mathcal{R}_{\mathrm{v} \alpha}}^2}{ E_{8\alpha} - E_{8\mathrm{v}}}
		\Bigg ],\\
		\gamma_3 =& \frac{2 m_0}{ \hbar^2} \sum_{\alpha  \neq \mathrm{v}} \Bigg [  \frac{1}{6} \frac{\abs{\mathcal{P}_{\mathrm{v} \alpha}}^2}{  E_{6\alpha} - E_{8\mathrm{v}}} +  \frac{1}{6} \frac{\abs{\mathcal{Q}_{\mathrm{v} \alpha}}^2}{  E_{7\alpha} - E_{8\mathrm{v}}}  \\
		& - \frac{1}{3} \frac{\abs{\mathcal{R}_{\mathrm{v} \alpha}}^2}{ E_{8\alpha} - E_{8\mathrm{v}}}
		\Bigg ].
	\end{align*}
	For the set of nonzero parameters considered here, the explicit form is
	\begin{align}
		\gamma_1 =& -1 + \frac{2 m_0}{ \hbar^2} \Bigg [  \frac{1}{3} \frac{\abs{{P}_{0}}^2}{E_{6\mathrm{c}} - E_{8\mathrm{v}} } + \frac{1}{3} \frac{\abs{{P}_{2}}^2}{E_{6\mathrm{q}} - E_{8\mathrm{v}} }  \nonumber
		\\ \nonumber & + \frac{1}{3} \frac{\abs{{P}_{1}'}^2}{E_{6\mathrm{w}} - E_{8\mathrm{v}} } +  \frac{1}{3} \frac{\abs{{Q}_{0}}^2}{  E_{7\mathrm{c}} - E_{8\mathrm{v}}}  \\ & + \frac{1}{3} \frac{\abs{{Q}_{0}}^2}{E_{8\mathrm{c}} - E_{8\mathrm{v}}} + \frac{4}{3} \frac{\abs{{R}_{0}}^2}{ E_{8\mathrm{t}} - E_{8\mathrm{v}}}
		\Bigg ],\\
		\label{eq:paramLutt1}
		\gamma_2 =& \frac{2 m_0}{ \hbar^2} \Bigg [  \frac{1}{6} \frac{\abs{{P}_{0}}^2}{E_{6\mathrm{c}} - E_{8\mathrm{v}} } + \frac{1}{6} \frac{\abs{{P}_{2}}^2}{E_{6\mathrm{q}} - E_{8\mathrm{v}} }  \nonumber
		+ \frac{1}{6} \frac{\abs{{P}_{1}'}^2}{E_{6\mathrm{w}} - E_{8\mathrm{v}} }  \\  & - \frac{1}{6} \frac{\abs{{Q}_{0}}^2}{E_{7\mathrm{c}} - E_{8\mathrm{v}}} + \frac{2}{3} \frac{\abs{{R}_{0}}^2}{ E_{8\mathrm{t}} - E_{8\mathrm{v}}}
		\Bigg ],\\
		\label{eq:paramLutt3}
		\gamma_3 =& \frac{2 m_0}{ \hbar^2} \Bigg [  \frac{1}{6} \frac{\abs{{P}_{0}}^2}{E_{6\mathrm{c}} - E_{8\mathrm{v}} } + \frac{1}{6} \frac{\abs{{P}_{2}}^2}{E_{6\mathrm{q}} - E_{8\mathrm{v}} } 
		+ \frac{1}{6} \frac{\abs{{P}_{1}'}^2}{E_{6\mathrm{w}} - E_{8\mathrm{v}} }  \nonumber \\  & + \frac{1}{6} \frac{\abs{{Q}_{0}}^2}{E_{7\mathrm{c}} - E_{8\mathrm{v}}} - \frac{1}{3} \frac{\abs{{R}_{0}}^2}{ E_{8\mathrm{t}} - E_{8\mathrm{v}}}
		\Bigg ].
	\end{align}
	
	To obtain  the parameters suitable for the 8-band $\kp$ model, one need to subtract the contributions from the couplings already accounted for in the Hamiltonian~\cite{Winkler2003}
	\begin{align*}
		\frac{m_0}{m'} =& \,  \frac{m_0}{m^*} - \frac{2 m_0}{\hbar^2} \qty[ \frac{2}{3} \frac{\abs{P_{0}}^2 }{E_{6\mathrm{c}} - E_{8\mathrm{v}}} + \frac{1}{3}  \frac{\abs{P_{0}}^2 }{E_{6\mathrm{c}} - E_{7\mathrm{v}}} ], \\
		\gamma'_1 =& \, \gamma_1 - \frac{2 m_0}{ \hbar^2}   \frac{1}{3} \frac{\abs{{P}_{0}}^2}{E_{6\mathrm{c}} - E_{8\mathrm{v}} },\\
		\gamma'_2 =& \, \gamma_2 - \frac{2 m_0}{ \hbar^2}   \frac{1}{6} \frac{\abs{{P}_{0}}^2}{E_{6\mathrm{c}} - E_{8\mathrm{v}} },\\
		\gamma'_3 =& \, \gamma_3 - \frac{2 m_0}{ \hbar^2}   \frac{1}{6} \frac{\abs{{P}_{0}}^2}{E_{6\mathrm{c}} - E_{8\mathrm{v}} }.
	\end{align*}
	
	The transition from the $8$- to $14$-band model involves further modification of the parameters by removing the couplings to ``${7\mathrm{c}}$" and ``${8\mathrm{c}}$"~\cite{Winkler2003}
	\begin{align*}
		\frac{m_0}{m''} =& \,  \frac{m_0}{m'} - \frac{2 m_0}{\hbar^2} \qty[ \frac{2}{3} \frac{\abs{P'_{0}}^2 }{E_{6\mathrm{c}} - E_{8\mathrm{c}}} + \frac{1}{3}  \frac{\abs{P'_{0}}^2 }{E_{6\mathrm{c}} - E_{7\mathrm{c}}} ], \\
		\gamma''_1 =& \, \gamma'_1 - \frac{2 m_0}{ \hbar^2} \qty[   \frac{1}{3} \frac{\abs{{Q}_{0}}^2}{  E_{7\mathrm{c}} - E_{8\mathrm{v}}}  + \frac{1}{3} \frac{\abs{{Q}_{0}}^2}{E_{8\mathrm{c}} - E_{8\mathrm{v}}} ],\\
		\gamma''_2 =& \, \gamma'_2 + \frac{2 m_0}{ \hbar^2} \frac{1}{6} \frac{\abs{{Q}_{0}}^2}{E_{7\mathrm{c}} - E_{8\mathrm{v}}},\\
		\gamma''_3 =& \, \gamma'_3 - \frac{2 m_0}{ \hbar^2} \frac{1}{6} \frac{\abs{{Q}_{0}}^2}{E_{7\mathrm{c}} - E_{8\mathrm{v}}}.
	\end{align*}

	Table \ref{tab:EPG123ME} contains the optical energy parameter ($E_\mathrm{P_0} = \frac{2m_0}{\hbar^2}|P_0|^2$), the Luttinger parameters ($\gamma_1$, $\gamma_2$, and $\gamma_3$), and the electron effective mass ($m^*$) for $\Gamma_{6c}$, for all the considered binary compounds.
	They were calculated using Eqs.~\ref{eq:paramEM}--\ref{eq:paramLutt3} with the parameters listed in Tables~\ref{tab:BX}-\ref{tab:InX}.
	\begin{table}[htbp]
		\centering
		\caption{The values of $E_{P_0}$, the Luttinger parameters, and the electron effective mass at the $\Gamma$ point in the band $\Gamma_{6c}$.
		}
		\begin{ruledtabular}
			\begin{tabular}{cccccc}
				& $E_{P_0}$(eV) & $\gamma_{1}$ & $\gamma_{2}$ & $\gamma_{3}$ & m$^*$ (m$_0$) \\
				\hline
				BN
				& 12.398 & 2.048 & 0.036 & 0.581 & 0.289$^a$ \\
				BP
				& 22.735 & 3.901 & -0.090 & 1.113 & 0.287$^a$ \\
				BAs
				& 22.877 & 4.685 & 0.107 & 1.443 & 0.204$^a$ \\
				BSb
				& 19.147 & 5.443 & 0.289 & 1.814 & 0.163$^b$ \\
				AlN
				& 17.782 & 1.559 & 0.392 & 0.613 & 0.274 \\
				AlP
				& 19.281 & 2.968 & 0.491 & 1.081 & 0.190 \\
				AlAs
				& 20.655 & 3.977 & 0.872 & 1.535 & 0.131 \\
				AlSb
				& 20.095 & 5.352 & 1.170 & 2.046 & 0.106 \\
				GaN
				& 14.807 & 2.631 & 0.671 & 1.012 & 0.191 \\
				GaP
				& 20.809 & 4.491 & 0.888 & 1.666 & 0.124 \\
				GaAs
				& 22.911 & 7.257 & 2.177 & 3.016 & 0.066 \\
				GaSb
				& 22.691 & 12.210 & 4.161 & 5.316 & 0.041 \\
				InN
				& 11.558 & 7.409 & 3.094 & 3.393 & 0.052 \\
				InP
				& 16.435 & 5.773 & 1.654 & 2.369 & 0.082 \\
				InAs
				& 18.493 & 16.882 & 7.102 & 7.891 & 0.026 \\
				InSb
				& 19.200 & 29.836 & 13.173 & 14.219 & 0.016
			\end{tabular}
		\end{ruledtabular}
		\begin{flushleft}
			\footnotesize{$^a$ third conduction band,\\ $^b$ second conduction band}
		\end{flushleft}
		\label{tab:EPG123ME}
	\end{table}
	
	Table \ref{tab:EGMASS} contains: the lattice parameters ($a_{\mathrm{lc}}$); the direct energy gaps at the $\Gamma$ point ($E_\mathrm{g}^\Gamma$); the indirect energy gaps to the $\Delta$ and $\Lambda$ valleys ($E_\mathrm{g}^{\Delta}$ and $E_\mathrm{g}^{\Lambda}$); the spin-orbit splittings of the top valence bands ($\Delta_{\mathrm{so}}$); the effective masses in the first conduction band at the $\Gamma$ point ($m_\mathrm{e}$); the absolute values of the effective masses of heavy holes and light holes along three crystallographic directions ($|m_{\mathrm{hh}}^{100}|$, $|m_{\mathrm{hh}}^{110}|$, $|m_{\mathrm{hh}}^{111}|$, $|m_{\mathrm{lh}}^{100}|$, $|m_{\mathrm{lh}}^{110}|$, and $|m_{\mathrm{lh}}^{111}|$); the effective masses in the split-off band at the $\Gamma$ point ($m_{\mathrm{so}}$); and the longitudinal effective masses in the $\Delta$- and $L$-valleys ($m_\Delta$ and $m_L$).
	All the lattice parameters are obtained as described in Sec.~\ref{sec:dft}.
	The remaining parameters are obtained numerically from the electronic-band structures computed within the 30-band $\kp$ with the parameters listed in Tables~\ref{tab:BX}-\ref{tab:InX}. 
	We note that $\Delta_{\mathrm{so}}$ are slightly different compared to the $\Delta_\mathrm{v}$ parameters presented in the main text. The reason is the influence of the off-diagonal spin-orbit coupling terms with parameter $\Delta^-$ which shifts the bands~\cite{Cardona1988}. While for most of the compounds its influence is negligible, in some cases (like BSb), the $\Delta_{\mathrm{so}}$ differs to the value of $\Delta_\mathrm{v}$ by a few percent.

	Effective masses at the $\Gamma$ point and at the minima of the $\Delta$- and $L$-valleys are evaluated with use of the standard formula
	\begin{equation*}
		\begin{aligned}
			m_n^{ijk}=\left[\frac{m_0}{\hbar^2}\frac{\partial^2E_n}{\partial k_{ijk}^2}\right]^{-1},
		\end{aligned}
	\end{equation*}
	where $ijk$ denotes chosen crystallographic direction and $n$ is a label of a given band.
	
	\renewcommand{\arraystretch}{1.2}
	\begin{table*}[t]
		\scriptsize
		\caption{The energy gaps, the spin-orbit splittings, and the effective masses calculated numerically based on the band structures obtained within the 30-band $\kp$ model. All the effective masses of holes are negative, so only the absolute values are listed in this table. We advise to examine band structures included in the Supplementary Material~\cite{Supplementary}.}
		\begin{ruledtabular}
			\begin{tabular}{llllllllllllllll}
				& $a_{\mathrm{lc}}$(\AA) & $E_\mathrm{g}^\Gamma$(eV) & $E_\mathrm{g}^\Delta$(eV) & $E_\mathrm{g}^\Lambda$(eV) & $\Delta_{\mathrm{so}}$(eV) & $m_\mathrm{e}$ & $|m_{\mathrm{hh}}^{100}|$ & $|m_{\mathrm{hh}}^{110}|$ & $|m_{\mathrm{hh}}^{111}|$ & $|m_{\mathrm{lh}}^{100}|$ & $|m_{\mathrm{lh}}^{110}|$ & $|m_{\mathrm{lh}}^{111}|$ & $|m_{\mathrm{so}}|$ & $m_\Delta$ & $\;\;\:m_\Lambda$\\
				\hline
				BN
				& 3.6061 & 11.214 & 6.595$^X$ & 12.593$^L$ & 0.024 & 1.072$^{7\mathrm{c}}$ & 0.517 & 0.964 & 1.130 & 0.472 & 0.332 & 0.316 & 0.480 & 1.170$^X$ & $\;\;\:$1.381$^L$ \\
				BP
				& 4.5227 & 4.289 & 1.913 & 4.768 & 0.046 & 0.358$^{7\mathrm{c}}$ & 0.269 & 0.508 & 0.597 & 0.249 & 0.173 & 0.165 & 0.257 & 1.150 & $\;\;\:$2.846 \\
				BAs
				& 4.7713 & 3.731 & 1.571 & 3.372 & 0.230 & 0.352$^{7\mathrm{c}}$ & 0.233 & 0.461 & 0.557 & 0.204 & 0.142 & 0.135 & 0.223 & 0.985 & $\;\;\:$2.081 \\
				BSb
				& 5.2250 & 1.224 & 1.113 & 2.454 & 0.379 & 0.359$^{7\mathrm{c}}$ & 0.233 & 0.449 & 0.553 & 0.167 & 0.124 & 0.118 & 0.199 & 0.960 & $\;\;\:$2.087 \\
				AlN
				& 4.3829 & 6.167 & 5.257$^X$ & 9.600$^L$ & 0.022 & 0.274 & 1.290 & 2.343 & 3.008 & 0.427 & 0.372 & 0.359 & 0.645 & 0.694$^X$ & $\;\;\:$1.268$^L$ \\
				AlP
				& 5.4719 & 4.406 & 2.534$^X$ & 3.907 & 0.066 & 0.190 & 0.504 & 0.969 & 1.240 & 0.253 & 0.204 & 0.195 & 0.343 & 1.041$^X$ & $\;\;\:$1.207 \\
				AlAs
				& 5.6764 & 2.983 & 2.251$^X$ & 3.050 & 0.324 & 0.131 & 0.451 & 0.851 & 1.103 & 0.175 & 0.148 & 0.142 & 0.277 & 1.118$^X$ & $\;\;\:$1.188 \\
				AlSb
				& 6.1578 & 2.179 & 1.634 & 1.839$^L$ & 0.658 & 0.106 & 0.341 & 0.623 & 0.795 & 0.130 & 0.111 & 0.107 & 0.237 & 1.574 & $\;\;\:$1.354$^L$ \\
				GaN
				& 4.5041 & 3.297 & 4.975 & 6.315$^L$ & 0.033 & 0.191 & 0.778 & 1.327 & 1.647 & 0.252 & 0.222 & 0.215 & 0.386 & 0.802 & $-$9.629$^L$ \\
				GaP
				& 5.4410 & 2.907 & 2.265 & 2.585 & 0.100 & 0.124 & 0.369 & 0.680 & 0.862 & 0.160 & 0.133 & 0.128 & 0.231 & 0.933 & $\;\;\:$1.454 \\
				GaAs
				& 5.6635 & 1.514 & 2.184 & 1.911$^L$ & 0.378 & 0.066 & 0.345 & 0.626 & 0.816 & 0.086 & 0.077 & 0.075 & 0.167 & 1.110 & $\;\;\:$1.437$^L$ \\
				GaSb
				& 6.1131 & 0.814 & 1.324$^X$ & 1.002$^L$ & 0.735 & 0.041 & 0.262 & 0.478 & 0.635 & 0.049 & 0.045 & 0.44 & 0.140 & 7.632$^X$ & $\;\;\:$1.589$^L$ \\
				InN
				& 4.9880 & 0.609 & 4.189$^X$ & 4.391$^L$ & 0.042 & 0.052 & 0.825 & 1.306 & 1.604 & 0.073 & 0.071 & 0.070 & 0.144 & 0.972$^X$ & $-$2.844$^L$ \\
				InP
				& 5.8810 & 1.423 & 2.355 & 2.210 & 0.125 & 0.082 & 0.419 & 0.750 & 0.967 & 0.110 & 0.099 & 0.096 & 0.186 & 0.933 & $\;\;\:$1.478 \\
				InAs
				& 6.0900 & 0.415 & 2.177 & 1.627 & 0.402 & 0.026 & 0.373 & 0.676 & 0.910 & 0.032 & 0.031 & 0.031 & 0.108 & 0.976 & $\;\;\:$1.441 \\
				InSb
				& 6.5191 & 0.235 & 1.588 & 0.919$^L$ & 0.762 & 0.016 & 0.287 & 0.525 & 0.716 & 0.018 & 0.017 & 0.017 & 0.120 & 1.204 & $\;\;\:$1.701$^L$ \\
			\end{tabular}
		\end{ruledtabular}
		\begin{flushleft}
			\footnotesize{$^X$ calculated exactly at the point X \\ $^L$ calculated exactly at the point L \\ $^{7\mathrm{c}}$ related to the band labeled as $\Gamma_{7\mathrm{c}}$}
		\end{flushleft}
		\label{tab:EGMASS}
	\end{table*}

	\section{Interband momentum matrix elements from the DFT}
	\label{app:mel}
	
	In this part, we describe a method to extract the  $\kp$ parameters directly from the DFT momentum matrix elements. The calculations were performed without the spin-orbit coupling.
	Although we worked with a reduced basis neglecting spin (15 states) in the numerical calculations, we present a derivation of the parameters with the full basis that includes spin (30 states) for the sake of consistency with the main part of the paper.
	
	
	The bare interband parameters are given with respect to the basis states (with modified and combined notations of Refs.~\cite{Winkler2003,Richard2004}) as
	\begin{align*}
		\mathcal{P}_{a b} &=  \frac{\hbar}{m_0}  \mel{S_a}{\hat{p}_x}{X_b},\\
		\mathcal{Q}_{a b} &=  \frac{\hbar}{m_0}  \mel{X_a}{\hat{p}_y}{Z_b},\\
		\mathcal{R}_{\mathrm{t} b}  &=  \frac{\hbar}{m_0}   \mel{D_{\mathrm{t},1}}{\hat{p}_x}{X_b} \\[2pt]
		& = -\frac{1}{\sqrt{3}}  \frac{\hbar}{m_0}  \mel{D_{\mathrm{t},2}}{\hat{p}_x}{X_b},
	\end{align*}
	where $a, b$ label the band sets: ``w", ``v", ``c", ``u", ``d", ``q" (except to ``t"). 
	%
	To obtain the value of $P_0 \equiv \mathcal{P}_{\mathrm{c} \mathrm{v}} =  \frac{\hbar}{m_0} \mel{S_\mathrm{c}}{\hat{p}_x}{X_\mathrm{v}}$ directly from the DFT, one need to calculate the elements $(p_{x})_{nm} = \mel{\Psi^\mathrm{(c)}_n}{\hat{p}_{x}}{\Psi^\mathrm{(v)}_m}$ between the states from the twofold degenerated conduction band ``6c" and the sixfold degenerated valence band ``8v $\oplus$ 7v" (as the spin-orbit coupling is neglected here).
	Hence the relevant block $(p_{x})_{nm}$ is $2 \times 6$. 
	The states can be further expressed as
	\begin{align*}
		\ket{\Psi^\mathrm{(c)}_n} &= \sum_{\sigma = \uparrow,\downarrow } \alpha^{(n)}_\sigma \ket{ S_\mathrm{c}} \otimes \ket{\sigma}, \\[2pt]
		\ket{\Psi^\mathrm{(v)}_m} &= \sum_{\sigma = \uparrow,\downarrow } \sum_{i=x,y,z}  \beta^{(m)}_{i \sigma} \ket{r_{\mathrm{v},i}} \otimes \ket{\sigma},
	\end{align*}
	where $\alpha^{(n)}_\sigma$, $\beta^{(m)}_{i \sigma}$ are complex coefficients, $\sigma$ denotes spin, and we use the notation $\ket{r_{\mathrm{v},x}} \equiv \ket{X_{\mathrm{v}}}$, $\ket{r_{\mathrm{v},y}} \equiv \ket{Y_{\mathrm{v}}}$, $\ket{r_{\mathrm{v},z}} \equiv \ket{Z_{\mathrm{v}}}$. The substitution leads to
	$$
	(p_{x})_{nm} =  \sum_{\sigma = \uparrow,\downarrow } \sum_{i=x,y,z} \alpha^{(n)*}_\sigma \beta^{(m)}_{i \sigma} \mel{S_\mathrm{c}}{\hat{p}_x}{r_{\mathrm{v},i}}.
	$$
	Taking a sum over the moduli square, one can obtain
	\begin{equation*}
		\sum_{n,m} \abs{(p_{x})_{nm} }^2 =  \sum_{\sigma = \uparrow,\downarrow } \sum_{i=x,y,z} \abs{ \mel{S_\mathrm{c}}{\hat{p}_x}{r_{\mathrm{v},i}}}^2.
	\end{equation*}
	Since $\mel{S_\mathrm{c}}{\hat{p}_x}{r_{\mathrm{v},i}} = \frac{m_0}{\hbar} P_0 \, \delta_{xi}$ and it does not depend on spin, one can get
	$$
	\abs{P_0} = \frac{\hbar}{m_0}  \sqrt{\frac{1}{2} \sum_{n,m} \abs{(p_{x})_{nm} }^2}.
	$$
	In the same way, one can find the other elements of the $\mathcal{P}_{ab}$ type.
	
	To calculate $Q_0 \equiv \mathcal{Q}_{\mathrm{c}\mathrm{v}} =  \frac{\hbar}{m_0} \mel{X_\mathrm{c}}{\hat{p}_y}{Z_\mathrm{v}}$ we take
	$$
	(p_{y})_{nm} = \sum_{\sigma} \sum_{i,j} \beta^{(n)*}_{i \sigma} \beta^{(m)}_{j \sigma} \mel{r_{\mathrm{c},i}}{\hat{p}_y}{r_{\mathrm{v},j}},
	$$
	where the block $(p_{y})_{nm}$ is $6 \times 6$ now.
	In the next step, one can obtain
	\begin{equation*}
		\sum_{n,m} \abs{(p_{y})_{nm} }^2 = 2 \sum_{i,j} \abs{ \mel{r_{\mathrm{c},i}}{\hat{p}_y}{r_{\mathrm{v},j}}}^2,
	\end{equation*}
	where the factor of 2 results from the summation over spin projections. Since $\abs{\mel{X_\mathrm{c}}{\hat{p}_y}{Z_\mathrm{v}}}^2 = \abs{\mel{Z_\mathrm{c}}{\hat{p}_y}{X_\mathrm{v}}}^2 = \frac{m^2_0}{\hbar^2} \abs{Q_0}^2$, one can get
	$$
	\abs{Q_0} = \frac{\hbar}{2 m_0} \sqrt{\sum_{n,m} \abs{(p_{y})_{nm} }^2}.
	$$
	The parameter $Q_1$ is calculated analogously.
	
	In the case of $R_0 = \mathcal{R}_{\mathrm{t}\mathrm{v}}$, using $(p_{x})_{nm} = \mel{\Psi^\mathrm{(t)}_n}{\hat{p}_{x}}{\Psi^\mathrm{(v)}_m}$, one needs to express
	\begin{align*}
		\ket{\Psi^\mathrm{(t)}_n} &= \sum_{\sigma = \uparrow,\downarrow } \sum_{i=1}^2 \gamma^{(n)}_{i,\sigma} \ket{D_{\mathrm{t},i}} \otimes \ket{\sigma}, \\[2pt]
	\end{align*}
	giving
	\begin{equation*}
		\sum_{n,m} \abs{(p_{x})_{nm} }^2 = 2 \sum_{i=1}^2 \sum_{j=x,y,z} \abs{ \mel{D_{\mathrm{t},i}}{\hat{p}_x}{r_{\mathrm{v},j}}}^2.
	\end{equation*}
	The non-zero contributions at the right-hand side come from  $\mel{D_{\mathrm{t},1}}{\hat{p}_x}{X_{\mathrm{v}}} = \frac{m_0}{\hbar} R_0$ and $\mel{D_{\mathrm{t},2}}{\hat{p}_x}{X_{\mathrm{v}}} = - \sqrt{3}\frac{m_0}{\hbar} R_0$.
	Therefore
	$$
	\abs{R_0} = \frac{\hbar}{ m_0} \sqrt{ \frac{1}{8} \sum_{n,m} \abs{(p_{x})_{nm} }^2}.
	$$
	The parameter $R_1$ can be obtained analogously.
	
	
	\bibliographystyle{prsty}
	\bibliography{abbr,./library.bib,MartaGlad.bib,Ref_DFT.bib,KG.bib,HSM.bib}
	
\end{document}